\documentclass[a4paper,11pt]{article}

\usepackage[margin=1.5cm,bottom=0.75cm,includefoot]{geometry}
\usepackage{graphicx}
\usepackage{mathtools}
\usepackage{amsmath}
\usepackage{amssymb}
\usepackage[colorlinks=true, allcolors=blue]{hyperref}
\usepackage{esint}
\usepackage[dvipsnames,table]{xcolor}
\usepackage[square,numbers,sort&compress]{natbib}
\usepackage{tikz}
\usetikzlibrary{3d,calc}
\usetikzlibrary{shadings}
\usetikzlibrary{arrows.meta}
\usepackage{caption}
\usepackage{placeins}
\usepackage{afterpage}
\usepackage{pgfmath}
\usepackage{pgfplots}
\usepgfplotslibrary{colormaps}
\usepackage{tikz-3dplot}
\usepackage{subcaption}
\usepackage{adjustbox}
\usepackage{array}
\usepackage{multirow}
\usepackage{makecell}
\usepackage{hhline}
\definecolor{LightGray}{rgb}{0.93,0.93,0.93}

\newcommand{\algbox}[1]{\vspace{0.0cm}\fcolorbox{LightGray}{LightGray}{\parbox{0.99\textwidth}{\color{black}\vspace{-0.2cm}#1 \vspace{-0.2cm}}}\vspace{0.2cm}}
\newtheorem{algorithm}{Algorithm}[section]
 
\newcommand{\comment}[1]{\textcolor{Gray}{\% #1}}
\renewcommand{\vec}[1]{\textbf{#1}}
\newcommand{\diff}{\text{d}}
\newcommand{\T}{\mathsf{T}}

\title{Functionally-graded drug delivery systems with binding reactions: analytical and stochastic approaches for the fraction of drug released\\}
\author{Obi A. Carwood\footnote{Corresponding Author (\href{obi.carwood@hdr.qut.edu.au}{obi.carwood@hdr.qut.edu.au})}\hspace{0.35em} and Elliot J. Carr\\[1ex] \small School of Mathematical Sciences, Queensland University of Technology (QUT), Brisbane, Australia}
\date{}

\begin{document}

\maketitle

\begin{abstract}
\noindent Mathematical modelling and computer simulation are increasingly being used alongside experiments to help optimise and guide the design of drug delivery systems. Recent drug delivery research has (i) highlighted the advantages of drug delivery systems constructed using functionally graded materials to achieve target release rates and desired dosage levels over time; and (ii) revealed how it is possible for drug to bind to the carrier material and become irreversibly immobilised within the system, reducing the amount of drug delivered. In this paper, we consider the effect of functionally graded materials and binding reactions on drug release from common slab, cylinder and sphere devices. In particular, two key contributions are presented. First, we outline a deterministic-continuum approach that develops exact analytical expressions for calculating the total fraction of drug released from the device based on a partial differential equation model of the release process. Second, we develop a stochastic-discrete approach for calculating the fraction of drug released over time based on a random-walk model  that captures the randomness of the release process and resulting variability in the total fraction of drug released. Both approaches are numerically validated and provide tools for exploring how the fraction of drug released depends on system parameters (e.g. diffusivity and reaction rate functions induced by the functionally graded material and binding reactions), insight which may be useful for designers of drug delivery systems.
\end{abstract}

\section{Introduction}

\begin{figure}[t]
\centering
\def\figh{0.35\textwidth}
\includegraphics[height=\figh]{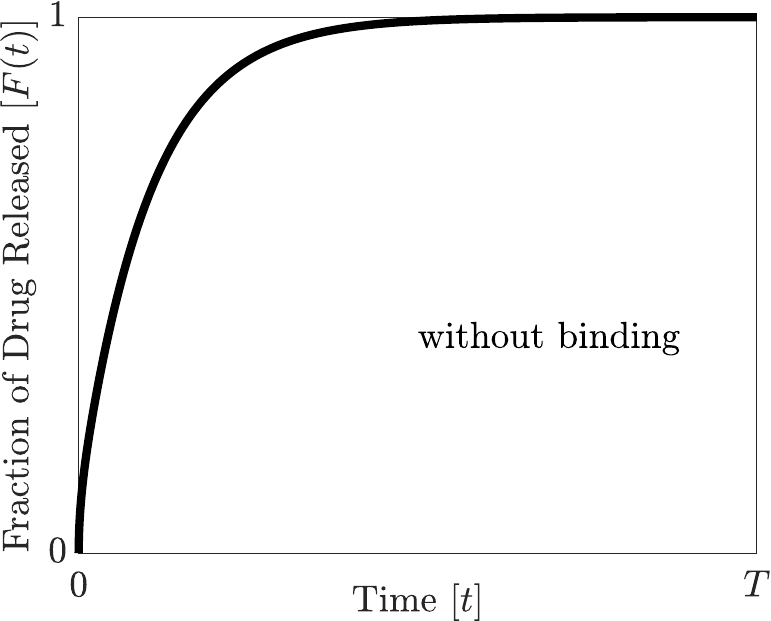}\hspace{1cm}\includegraphics[height=\figh]{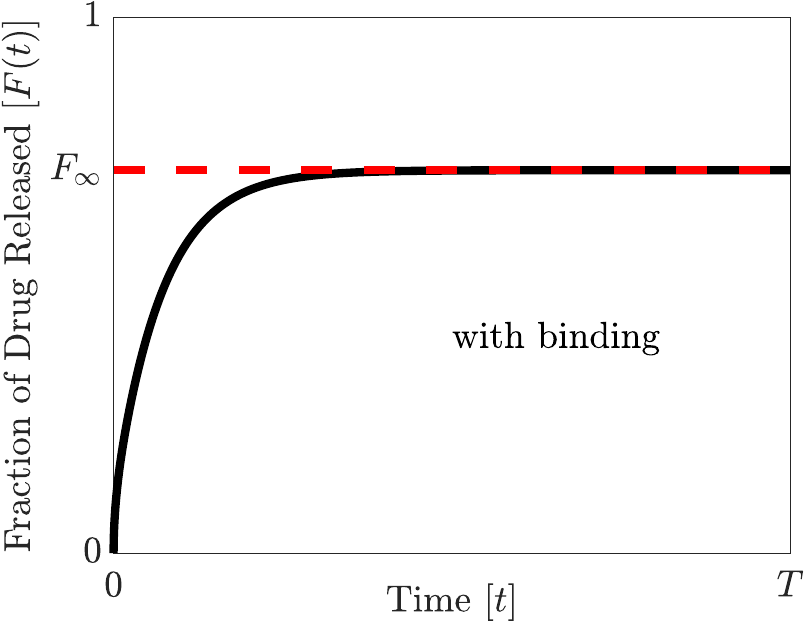}
\caption{\textbf{Drug release profiles without/with binding reactions.} Typical drug release profiles without binding reactions and with binding reactions, where $F(t)$ is the amount of drug released up to time $t$ scaled by the initial amount of drug present in the device and $\smash{F_{\infty} := \lim_{t\rightarrow\infty}F(t)}$ is the fraction of drug released. Note that $\smash{F_{\infty} = 1}$ (all drug is released) without binding and $\smash{F_{\infty} < 1}$ (some drug is never released) with binding.}
\label{fig:release_profiles}
\end{figure}

\begin{figure}[h]
    \centering
\includegraphics[width=0.99\textwidth]{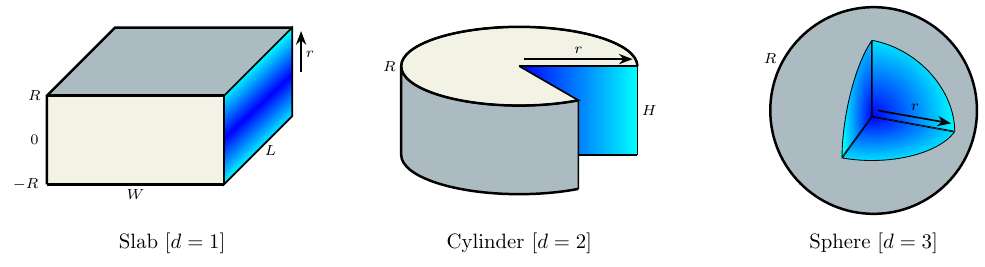}
    \caption{\textbf{Functionally graded material (FGM) delivery systems.} Radially-symmetric FGM delivery devices of slab, cylinder and sphere geometry, all with radius $R$. The colour gradient highlights the continuously varying material properties (e.g. diffusivity and reaction-rate, see equations (\ref{PDE: Governing Reaction-Diffusion})--(\ref{EQS: Governing BC2})) of the FGM in the direction of the radial coordinate $r$. For the slab geometry, $L\gg R$ and $W \gg R$, such that the drug release is assumed to be negligible through the minor surfaces of area $2RL$ and $2RW$ relative to the major surfaces of area $LW$. Similarly, for the cylinder geometry, $H \gg R$, such that the drug release is assumed to be negligible through the minor surfaces of areas $\pi R^2$ relative to the major surface of area $2\pi RH$. Slab and cylinder geometries not to scale. Shading (\textcolor[RGB]{172 187 194}{\rule{1.2em}{0.7em}}) indicates the release surfaces and thin coating, which is either fully-permeable or semi-permeable.}
    \label{FIG: Device Geometries}
\end{figure}

Modelling and simulation are increasingly becoming powerful tools in drug delivery research. Mathematical models improve the fundamental understanding of drug transport and release, provide insights into the effect of device design parameters on drug release profiles, reduce the need for costly and time-consuming experiments, and help optimise and guide the design of drug delivery systems to achieve desired drug release profiles \cite{Sahai:2021,Siepmann:2008,Lotter:2023,Montelongo:2022}.

Mathematical models of drug delivery can usually be described as either \textit{deterministic-continuum} or \textit{stochastic-discrete}. Deterministic-continuum models assume the distribution of drug within the device is described by the drug concentration, a continuous function of space changing over time, with a deterministic partial differential equation and appropriate boundary conditions governing drug movement and release, respectively \cite{Siepmann:2012,Carr:2018_Layered,Kaoui:2018,Arifin:2006,Gonçalves:2023}. In contrast, stochastic-discrete models assume the distribution of drug within the device is described by the individual positions of drug particles, discrete points in space changing over time, with probabilities governing drug movement and release events \cite{Hadjitheodorou:2013,Gomes:2020,Filippini:2023,Ignacio:2022,Carr:2022_Surrogate}. Deterministic-continuum models do not account for randomness, may be amendable to analytical solution, and rely on the assumption of a large number of drug particles. On the other hand, stochastic-discrete models account for variability and randomness in the drug release process, hold for small and large numbers of drug particles, and may be computationally prohibitive for a large number of drug particles. Due to these differences, it is important to study both deterministic-continuum and stochastic-discrete models of drug delivery.

Deterministic-continuum models of drug delivery are well established and are typically based on Fick’s second law \cite{Carr:2018_Layered,Carr:2024_FunctionalRelease,Ignacio:2021,Gomes:2020,Arifin:2006,Siepmann:2008,Hu:2024}, where the spatial and temporal behaviour of the drug
concentration is governed by the diffusion equation and specified initial and boundary conditions. Diffusion-only models of this kind yield a drug release profile $F(t)$ (amount of drug released up to time $t$ scaled by the initial amount of drug present in the device) that tends to one in the long time limit (Figure \ref{fig:release_profiles}). Recent experimental work \cite{Toniolo:2018}, however, has suggested that some drug particles are never released, likely due to undergoing a chemical reaction and binding to the carrier material \cite{Jain:2022,Pontrelli:2021,Jain:2024}. This binding mechanism can be captured in deterministic-continuum models by including a first-order reaction term that irreversibly immobilises drug at a specified rate  \cite{Jain:2022,Pontrelli:2021,Carr:2024_FunctionalRelease,Carr:2024_FractionRelease}. In this case, resulting reaction-diffusion models yield a drug release profile $F(t)$ that tends (in the long time limit) to a value, $F_{\infty}$, which is less than one  \cite{Carr:2024_FractionRelease,Carr:2024_FunctionalRelease} (Figure \ref{fig:release_profiles}).

Both deterministic-continuum and stochastic-discrete models of drug delivery are commonly limited to either homogeneous or composite carriers with constant or piecewise-constant material properties (e.g. diffusivity), respectively \cite{Siepmann:2012,Carr:2018_Layered,Kaoui:2018,Jain:2022,Mubarak:2021,Filippini:2023,Ignacio:2022,Carr:2022_Surrogate,Hadjitheodorou:2013}. Recently, however, there has been an increase in focus on functionally graded materials (FGMs), which exhibit smooth and continuously varying material properties. The implementation of FGMs have proven to be a promising way to control the drug release, owing to their spatially varying properties, which enable improved control of diffusion kinetics \cite{Saleh:2020,Shinohara:2013,Bretti:2023}. Moreover, the architecture of FGMs allow for drug carrier designs to target specific release rates which offers significant advantages over conventional homogeneous or composite systems \cite{Bretti:2023,Carr:2024_FunctionalRelease}.

Several recent studies have focussed on binding reactions and/or functional graded materials in the context of deterministic-continuum models of drug delivery. First, \cite{Bretti:2023} proposed a diffusion model for an FGM system of slab geometry without binding reactions, outlining a numerical solution for exploring the drug concentration and drug release profiles. Next, \cite{Carr:2024_FunctionalRelease} considered a reaction-diffusion model for a spherical FGM system with binding reactions, developing a semi-analytical solution for the drug concentration and drug release profiles. Finally, \cite{Carr:2024_FractionRelease} considered a reaction-diffusion model for homogeneous (monolithic) and composite (core-shell) systems of slab, cylinder and sphere geometry with binding reactions, developing closed-form expressions for the fraction of drug released ($F_{\infty}$).

In the current paper, we build on and extend this recent work, presenting two key advances for modelling drug release from functionally-graded delivery systems with binding reactions. First, we outline a deterministic-continuum approach that develops exact analytical expressions for calculating the total fraction of drug released, $F_{\infty}$. Second, we develop a stochastic-discrete approach for calculating the fraction of drug released over time, $F(t)$, that captures the randomness inherent in the drug delivery process and the resulting variability in the total fraction of drug released. Underpinning both approaches is the recently established reaction-diffusion model presented in \cite{Carr:2024_FunctionalRelease}, where smooth spatially-varying diffusivity and reaction rate functions accommodate the functionally-graded material and binding reactions. The deterministic-continuum approach shows how $F_{\infty}$ can be analytically formulated from the governing reaction-diffusion model and exactly resolved via an eigenfunction expansion solution of an appropriate boundary value problem. On the other hand, the stochastic-discrete approach develops a random-walk model of the drug release process by discretising the governing reaction-diffusion model in space and time to derive probabilities governing the movement, binding and release of individual drug particles. Both approaches are valid for radially-symmetric slab, cylinder and sphere devices encapsulated in a thin coating that may be fully-permeable (zero resistance to drug release) or semi-permeable (finite resistance to drug release) (see Figure \ref{FIG: Device Geometries}). In summary, the work provides analytical insight into, and computational tools for exploring, the effect of system parameters (diffusivity and reaction rate function induced by the functionally graded material, geometry of the device, coating permeability) on the fraction of drug released, insight which may be useful for designers and manufacturers of drug delivery systems.

The remaining sections of the paper are arranged as follows. Sections \ref{SEC: Analytical Approach} and \ref{SEC: Stochastic Approach} outline our deterministic-continuum and stochastic-discrete approaches, respectively, while section \ref{SEC: Results} provides numerical experiments validating and demonstrating application of both approaches. Finally, section \ref{SEC: Conclusions} concludes the work, and provides some possible directions for future research.

\section{Deterministic-Continuum Approach}
\label{SEC: Analytical Approach}

We now outline our deterministic-continuum approach for calculating the total fraction of drug released, $F_{\infty}$, for the functionally-graded drug delivery systems considered in this work.
\subsection{Reaction-diffusion model}

We assume that drug release is governed by the reaction-diffusion model presented in \cite{Carr:2024_FunctionalRelease}:
\begin{gather}
    \label{PDE: Governing Reaction-Diffusion}
    \frac{\partial c}{\partial t} = \frac{1}{r^{d-1}} \frac{\partial}{\partial r} \left( D(r) r^{d-1} \frac{\partial c}{\partial r} \right) - k(r)c, \quad \quad 0<r<R, \quad \quad 0 < t < T,\\
    \label{EQ: Governing IC}
    c(r,0) = c_0(r), \quad \quad 0 \leq r \leq R, \\
    \label{EQS: Governing BC1}
        \frac{\partial c}{\partial r}(0,t) = 0,\\
    \label{EQS: Governing BC2}
    \begin{dcases}
        c(R,t) = 0, \quad &\text{if fully-permeable,} \\
        -D(R) \frac{\partial c}{\partial r}(R,t) = Pc(R,t), \quad &\text{if semi-permeable,}
    \end{dcases}
\end{gather}
where $d=1,2,3$ for the slab, cylinder, sphere, respectively, $R$ is the radius of the device, $c(r,t)$ is the drug concentration at radius $r$ and time $t$, $T$ is a specified end time, $c_0(r)$ is the initial concentration profile, $P$ is the mass transfer coefficient characterising the permeability of the thin coating encapsulating the device at $r = R$ (see Figure~\ref{FIG: Device Geometries}), and $D(r)$ and $k(r)$ are the radially-dependent diffusivity and reaction-rate functions prescribed by the functionally-graded material. The boundary condition at $r=0$ (\ref{EQS: Governing BC1}) is the standard boundary condition for radially-symmetric geometries while the boundary condition at $r=R$ (\ref{EQS: Governing BC2}) describes the standard boundary conditions for fully-permeable or semi-permeable coatings~\cite{Kaoui:2018}.

\subsection{Total fraction of drug released}
The total fraction of drug released, $F_{\infty}$, is defined as the total cumulative amount of drug released from the device scaled by the initial amount of drug present in the device \cite{Carr:2024_FractionRelease}. For the radially-symmetric model (\ref{PDE: Governing Reaction-Diffusion})--(\ref{EQS: Governing BC2}), $F_{\infty}$ can be expressed as follows
\begin{align}
    \label{EQ: Finf - c}
    F_{\infty} = \frac{R^{d-1} \int_0^{\infty} -D(R) \frac{\partial c}{\partial r}(R,t) \, \diff t}{\int_0^R r^{d-1} c_0(r) \, \diff r},
\end{align}
which is a straightforward generalisation of the formula presented in \cite{Carr:2024_FractionRelease} to accommodate the radially-dependent parameters, $D(r)$, $k(r)$ and $c_0(r)$ (see \ref{APPENDIX: Derivation for Finf}). One of our primary goals in this paper is to develop analytical expressions that highlight how $F_{\infty}$ depends on the model parameters: diffusivity, $D(r)$, reaction-rate, $k(r)$, device radius, $R$, mass transfer coefficient, $P$, and initial concentration profile, $c_{0}(r)$. To achieve this, we follow the approach in \cite{Carr:2024_FractionRelease} and introduce the following transformation:
\begin{align}
    \label{EQ: Transformation c to C}
    C(r) = \int_0^{\infty} c(r,t) \, \diff t,
\end{align}
to derive an alternative form for the fraction of drug released. The relevant working, given in \ref{APPENDIX: New expression for Finf}, shows that $F_{\infty}$ can be reformulated in terms of $C(r)$:
\begin{align}
    \label{EQ: Finf - C}
    F_{\infty} = 1 - \frac{\int_0^R r^{d-1} k(r) C(r) \, \diff r}{\int_0^R r^{d-1} c_0(r) \, \diff r},
\end{align}
and that $C(r)$ satisfies the following boundary value problem:
\begin{gather}
    \label{ODE: Transformed Reaction-Diffusion - C}
    \frac{1}{r^{d-1}} \frac{\diff}{\diff r} \left( D(r) r^{d-1} \frac{\diff C}{\diff r} \right) - k(r)C = -c_0(r),\\
    \label{EQ: Transformed BC1 - C}
    \frac{\diff C}{\diff r}(0) = 0,\\
    \label{EQ: Transformed BC2 - C}
    \begin{dcases}
        C(R) = 0, \quad &\text{if fully-permeable,} \\
        -D(R)\frac{\diff C}{\diff r}(R) = PC(R), \quad &\text{if semi-permeable.}
    \end{dcases}
\end{gather}
Note that this reformulation avoids the full time-dependent solution for $c(r,t)$, which has a significantly more complicated analytical form than $C(r)$. 

\subsection{Non-dimensionalisation}
\noindent Before proceeding further, we non-dimensionalise the problem by introducing the following dimensionless variables,
\begin{align}
    \label{EQS: non-dimensionalised variables}
    \hat{r} := \frac{r}{R}, \quad \hat{C}(\hat{r}) := \frac{C(r)}{C^*}, \quad \hat{D}(\hat{r}):= \frac{D(r)}{D^*}, \quad \hat{k}(\hat{r}) := \frac{R^2}{D^*} k(r), \quad \hat{c}_0(\hat{r}) := \frac{R^2}{C^* D^*} c_0(r), \quad \hat{P} := \frac{PR}{D^*},
\end{align}
where 
\begin{align*}
C^* = \underset{r \in [0,R]}{\max} c_0(r) \times T,\quad D^* = \underset{r \in [0,R]}{\max} D(r).
\end{align*} 
Substituting the dimensionless variables (\ref{EQS: non-dimensionalised variables}) into (\ref{EQ: Finf - C})--(\ref{EQ: Transformed BC2 - C}) yields the following equivalent form for $F_{\infty}$ (\ref{EQ: Finf - C}):
\begin{align}
    \label{EQ: Finf - Chat}
    F_{\infty} = 1 - \frac{\int_0^1 \hat{r}^{d-1} \hat{k}(\hat{r}) \hat{C}(\hat{r}) \, \diff\hat{r}}{\int_0^1 \hat{r}^{d-1} \hat{c}_0(\hat{r}) \, \diff\hat{r}},
\end{align}
where $\hat{C}(\hat{r})$ satisfies the following non-dimensional form of the boundary value problem (\ref{ODE: Transformed Reaction-Diffusion - C})--(\ref{EQ: Transformed BC2 - C}),
\begin{gather}
    \label{ODE: Transformed Reaction-Diffusion - Chat}
    \frac{1}{\hat{r}^{d-1}} \frac{\diff}{\diff\hat{r}} \left(\hat{D}(\hat{r})\hat{r}^{d-1} \frac{\diff\hat{C}}{\diff\hat{r}} \right) - \hat{k}(\hat{r})\hat{C} = -\hat{c}_0(\hat{r}),\\
    \label{EQ: Transformed BC1 - Chat}
    \frac{\diff\hat{C}}{\diff\hat{r}}(0) = 0,\\
    \label{EQ: Transformed BC2 - Chat}
    \begin{dcases}
        \hat{C}(1) = 0, \quad &\text{if fully-permeable,} \\
        -\hat{D}(1)\frac{\diff\hat{C}}{\diff\hat{r}}(1) = \hat{P}\hat{C}(1), \quad &\text{if semi-permeable.}
    \end{dcases}
\end{gather}
While the non-dimensionalised system is only marginally simpler than the original system, the rescaling achieved through non-dimensionalising is important to ensure the magnitude of most model parameters (some of which are often very small \cite{Carr:2024_FractionRelease,Carr:2024_FunctionalRelease}) are close to one.

\subsection{Analytical solution}
\noindent To solve the boundary value problem (\ref{ODE: Transformed Reaction-Diffusion - Chat})--(\ref{EQ: Transformed BC2 - Chat}), we adapt previous approaches for dealing with spatially-dependent coefficients \cite{Carr:2024_FunctionalRelease,Johnston:1991,Cotta:2009,Liu:2000} and expand $\hat{C}(\hat{r})$ in an infinite series involving ``simple'' eigenvalues and eigenfunctions. Following \cite{Carr:2024_FunctionalRelease}, the eigenvalues $\{\lambda_{n}$, $n\in\mathbb{N}^{+}\}$ and eigenfunctions $\{X_{n}(\hat{r})$, $n\in\mathbb{N}^{+}\}$ are obtained from the following Sturm-Liouville problem:
\begin{gather}
    \label{ODE: Sturm-Liouville - X(r)}
    \frac{1}{{\hat{r}}^{d-1}} \frac{\text{d}}{\text{d} \hat{r}} \left( {\hat{r}}^{d-1} \frac{\text{d}X_{n}}{\text{d}\hat{r}} \right) = -\lambda_{n}^2X_{n},\\
    \label{ODE: Sturm-Liouville BC1}
    \frac{\text{d}X_n}{\text{d}\hat{r}}(0) = 0,\\
    \label{ODE: Sturm-Liouville BC2}
    \begin{dcases}
        X_n(1) = 0, \quad &\text{if fully-permeable,} \\
        -\hat{D}(1)\frac{\diff X_n}{\diff \hat{r}}(1) = \hat{P}X_n(1), \quad &\text{if semi-permeable,}
    \end{dcases}
\end{gather}
involving the radially-symmetric Laplace operator and the (already) homogeneous boundary conditions (\ref{EQ: Transformed BC1 - Chat})--(\ref{EQ: Transformed BC2 - Chat}).

Solving the Sturm-Liouville problem (\ref{ODE: Sturm-Liouville - X(r)})--(\ref{ODE: Sturm-Liouville BC2}), yields the following well-known sets of eigenvalues and eigenfunctions depending on the device geometry ($d=1,2,3$ for the slab, cylinder, sphere, respectively) and the coating permeability (fully-permeable or semi-permeable):

\bigskip\noindent\textit{Fully-permeable coating}:
\begin{align}
    \label{EQS: SL - Eigenvalues (fully-permeable)}
    \lambda_n &= \begin{dcases}
        \left( n - \tfrac{1}{2} \right) \pi, \quad &\text{if $d=1$}, \\
        \text{$n$th positive root of $J_0(\lambda) = 0$}, \quad &\text{if $d=2$}, \\
        n\pi, \quad &\text{if $d=3$},
    \end{dcases}
\end{align}
\begin{align}
    \label{EQS: SL - Eigenfunctions (FP)}
    X_n(\hat{r}) &=
    \begin{dcases}
        \sqrt{2} \cos(\lambda_n \hat{r}), \quad &\text{if $d=1$}, \\
        \frac{\sqrt{2}}{J_1(\lambda_n)} J_0(\lambda_n \hat{r}), \quad &\text{if $d=2$}, \\
        \frac{\sqrt{2} \sin(\lambda_n \hat{r})}{\hat{r}}, \quad &\text{if $d=3$},
    \end{dcases}
\end{align}
\noindent\textit{Semi-permeable coating}:
\begin{align}
    \label{EQS: SL - Eigenvalues (semi-permeable)}
    \lambda_{n} = 
    \begin{dcases}
        \text{$n$th positive root of $\lambda \hat{D}(1) \tan(\lambda) = \hat{P}$}, \quad &\text{if $d=1$}, \\
        \text{$n$th positive root of $\lambda \hat{D}(1) J_1(\lambda) = \hat{P} J_0(\lambda)$}, \quad &\text{if $d=2$}, \\
        \text{$n$th positive root of $(\hat{D}(1)-\hat{P})\tan(\lambda) = \lambda \hat{D}(1)$}, \quad &\text{if $d=3$},
    \end{dcases}
\end{align}
\begin{align}
    \label{EQS: SL - Eigenfunctions}
    X_n(\hat{r}) =
    \begin{dcases}
        \frac{2\sqrt{\lambda_n}}{\sqrt{\sin(2\lambda_n)+2\lambda_n}}\cos(\lambda_n \hat{r}), \quad &\text{if $d=1$}, \\
        \frac{\sqrt{2}}{\sqrt{J_0(\lambda_n)^2+J_1(\lambda_n)^2}}J_0(\lambda_n \hat{r}), \quad &\text{if $d=2$}, \\
        \frac{2\sqrt{\lambda_n}}{\sqrt{2\lambda_n - \sin(2\lambda_n)}} \frac{\sin(\lambda_n \hat{r})}{\hat{r}}, \quad &\text{if $d=3$},
    \end{dcases}
\end{align}
where $J_{0}$ and $J_{1}$ are, respectively, the order $0$ and $1$ Bessel functions of the first kind. Note that eigenfunctions are orthogonal and have been normalised to ensure they have length one, that is,
\begin{align}
    \label{EQ: orthonormal property - X}
    \int_0^1 \hat{r}^{d-1} X_n(\hat{r})X_m(\hat{r}) \, \diff \hat{r} = \begin{dcases} 1, & \text{if $m=n$,}\\ 0, & \text{if $m\neq n$.}\end{dcases}
\end{align}
With the eigenpairs $(\lambda_n,X_n)$ identified for all cases, the analytical solution for $\hat{C}(\hat{r})$ is expressed as
\begin{align}
    \label{EQ: Infinite Series - C(r)}
    \hat{C}(\hat{r}) = \sum_{n=1}^{\infty} \alpha_n X_n(\hat{r}),
\end{align}
where, due to orthonormality (\ref{EQ: orthonormal property - X}), we have
\begin{align}
    \label{EQ: orthonormal property - alpha}
    \alpha_n = \int_0^1 \hat{r}^{d-1} \hat{C}(\hat{r}) X_n(\hat{r}) \, \diff \hat{r}.
\end{align}
To complete the solution procedure, we calculate the coefficients $\{\alpha_{n}, n\in\mathbb{N}^{+}\}$ to ensure the solution expansion (\ref{EQ: Infinite Series - C(r)}) satisfies the non-dimensionalised ODE (\ref{ODE: Transformed Reaction-Diffusion - Chat}). This is achieved by substituting (\ref{EQ: Infinite Series - C(r)}) into (\ref{ODE: Transformed Reaction-Diffusion - Chat})
\begin{align*}
    \sum_{n=1}^{\infty} \alpha_n \left( \frac{1}{\hat{r}^{d-1}} \frac{\diff}{\diff \hat{r}} \left( \hat{D}(\hat{r}) \hat{r}^{d-1} \frac{\diff X_n}{\diff \hat{r}} \right) - \hat{k}(\hat{r}) X_n(\hat{r}) \right) = -\hat{c}_0(\hat{r}),
\end{align*}
applying the product rule of differentiation
\begin{align*}
    \sum_{n=1}^{\infty} \alpha_n \left( \hat{D}'(\hat{r}) {X'_n}(\hat{r}) + \hat{D}(\hat{r}) \frac{1}{\hat{r}^{d-1}} \frac{\diff}{\diff \hat{r}} \left( \hat{r}^{d-1} \frac{\diff X_n}{\diff \hat{r}} \right) - \hat{k}(\hat{r}) X_n(\hat{r}) \right) &= -\hat{c}_0(\hat{r}),
\end{align*}
utilising equation (\ref{ODE: Sturm-Liouville - X(r)})
\begin{align*}
    \sum_{n=1}^{\infty} \alpha_n \left( \hat{D}'(\hat{r}) {X'_n}(\hat{r}) - \left( {\lambda^{2}_n}\hat{D}(\hat{r}) + \hat{k}(\hat{r}) \right) X_n(\hat{r}) \right) &= -\hat{c}_0(\hat{r}),
\end{align*}
and, finally, multiplying by $\hat{r}^{d-1}X_m(\hat{r})$ and integrating over $\hat{r} \in [0,1]$:
\begin{align}
\label{EQ: linear equation - alpha}
    \sum_{n=1}^{\infty} \alpha_n \int_0^1 \hat{r}^{d-1}\left( \hat{D}'(\hat{r}) {X'_n}(\hat{r})X_m(\hat{r}) - \left( {\lambda_n^{2}}\hat{D}(\hat{r}) + \hat{k}(\hat{r}) \right) X_n(\hat{r}) X_m(\hat{r}) \right) \diff \hat{r} &= - \int_0^1 \hat{r}^{d-1} \hat{c}_0(\hat{r}) X_m(\hat{r}) \, \diff \hat{r}.
\end{align}
Note that equation (\ref{EQ: linear equation - alpha}) is linear in the coefficients $\{\alpha_{n}, n\in\mathbb{N}^{+}\}$ and hence assembling this equation for each $m\in\mathbb{N}^+$ yields a linear system
\begin{align}
\label{EQ: linear system - alpha}
\mathbf{A}\boldsymbol{\alpha} = \mathbf{b},
\end{align}
where $A_{m,n}$ (entry of $\mathbf{A}$ in row $m$ and column $n$) and $b_{m}$ ($m$th entry of $\mathbf{b}$) are defined as
\begin{align}
    \label{EQ: alpha linear system - A form}
    A_{m,n} &= \int_0^1 \hat{r}^{d-1}\left( \hat{D}'(\hat{r}) {X'_n}(\hat{r})X_m(\hat{r}) - \left(\lambda_n^2\hat{D}(\hat{r}) + \hat{k}(\hat{r}) \right) X_n(\hat{r}) X_m(\hat{r}) \right) \diff \hat{r},\\
    \label{EQ: alpha linear system - b form}
    b_m &= - \int_0^1 \hat{r}^{d-1} \hat{c}_0(\hat{r}) X_m(\hat{r}) \, \diff \hat{r}.
\end{align}
Solving this linear system for $\boldsymbol{\alpha} = (\alpha_{1},\alpha_{2},\cdots)^{\T}$ provides the coefficients in the solution expansion (\ref{EQ: Infinite Series - C(r)}), and thus the analytical solution for $\hat{C}(\hat{r})$ is fully resolved.

\subsection{Analytical expressions}
\noindent Substituting the analytical solution for $\hat{C}(\hat{r})$ (\ref{EQ: Infinite Series - C(r)}) into $F_{\infty}$ (\ref{EQ: Finf - Chat}) yields an analytical expression for the total fraction of drug released depending on the device geometry (slab, cylinder or sphere) and the coating permeability (fully-permeable or semi-permeable):

\bigskip\noindent\textit{Fully-permeable coating}:
\begin{alignat}{2}
    \label{EQ: Finf Final Result Fully - d=1}
    &\textit{Slab:} \qquad& F_{\infty} &= 1 - \sum_{n=1}^{\infty} \frac{\sqrt{2}\alpha_n}{\int_0^1 \hat{c}_0(\hat{r}) \,\text{d}\hat{r}} \int_0^1 \hat{k}(\hat{r}) \cos(\lambda_n \hat{r}) \,\text{d}\hat{r},\\
    \label{EQ: Finf Final Result Fully - d=2}
    &\textit{Cylinder:} \qquad& F_{\infty} &= 1 - \sum_{n=1}^{\infty} \frac{\sqrt{2}\alpha_n}{|\text{J}_1(\lambda_n)| \int_0^1 \hat{r} \hat{c}_0(\hat{r}) \,\text{d}\hat{r}} \int_0^1 {\hat{r}} \hat{k}(\hat{r}) \text{J}_0(\lambda_n \hat{r}) \,\text{d}\hat{r},\\
    \label{EQ: Finf Final Result Fully - d=3}
    &\textit{Sphere:} \qquad& F_{\infty} &= 1 - \sum_{n=1}^{\infty} \frac{\sqrt{2}\alpha_n}{\int_0^1 {\hat{r}}^2 \hat{c}_0(\hat{r}) \,\text{d}\hat{r}} \int_0^1 {\hat{r}} \hat{k}(\hat{r}) \sin(\lambda_n \hat{r}) \,\text{d}\hat{r},
\intertext{\textit{Semi-permeable coating}:}
    \label{EQ: Finf Final Result Semi - d=1}
    &\textit{Slab:} \qquad& F_{\infty} &= 1 - \sum_{n=1}^{\infty} \frac{2\sqrt{\lambda_n}\alpha_n}{\sqrt{\sin(2\lambda_n) + 2\lambda_n}\int_0^1 \hat{c}_0(\hat{r}) \,\text{d}\hat{r}} \int_0^1 \hat{k}(\hat{r}) \cos(\lambda_n \hat{r}) \,\text{d}\hat{r},\\
    \label{EQ: Finf Final Result Semi - d=2}
    &\textit{Cylinder:} \qquad& F_{\infty} &= 1 - \sum_{n=1}^{\infty} \frac{\sqrt{2}\alpha_n}{\sqrt{\text{J}_0(\lambda_n)^2 + \text{J}_1(\lambda_n)^2} \int_0^1 \hat{r} \hat{c}_0(\hat{r}) \,\text{d}\hat{r}} \int_0^1 {\hat{r}} \hat{k}(\hat{r}) \text{J}_0(\lambda_n \hat{r}) \,\text{d}\hat{r},\\
    \label{EQ: Finf Final Result Semi - d=3}
    &\textit{Sphere:} \qquad& F_{\infty} &= 1 - \sum_{n=1}^{\infty} \frac{2\sqrt{\lambda_n}\alpha_n}{\sqrt{2\lambda_n - \sin(2\lambda_n)}\int_0^1 {\hat{r}}^2 \hat{c}_0(\hat{r}) \,\text{d}\hat{r}} \int_0^1 {\hat{r}} \hat{k}(\hat{r}) \sin(\lambda_n \hat{r}) \,\text{d}\hat{r},
\end{alignat}
where the eigenvalues $\{\lambda_{n}$, $n\in\mathbb{N}^{+}\}$ are defined in equations (\ref{EQS: SL - Eigenvalues (fully-permeable)}) and (\ref{EQS: SL - Eigenvalues (semi-permeable)}),  the coefficients $\{\alpha_{n}$, $n\in\mathbb{N}^{+}\}$ are defined via solution of the linear system (\ref{EQ: linear system - alpha}), and the dimensionless variables are defined in equation (\ref{EQS: non-dimensionalised variables}). Finally, we note that in practice the analytical expressions (\ref{EQ: Finf Final Result Fully - d=1})--(\ref{EQ: Finf Final Result Semi - d=3}) are computed using a specified (finite) number of terms, $N_s$.

\section{Stochastic-Discrete Approach}
\label{SEC: Stochastic Approach}
We now outline our stochastic-discrete approach for determining the fraction of drug released based on a random-walk model of the drug delivery process.  

\subsection{Determination of probabilities}
Random walk models of diffusion can be developed by discretising the diffusion model in space and time to derive probabilities governing the movement of individual particles \cite{Carr:2024_FractionRelease,Meinecke:2016,Cai:2006,Anderson:1998}. In this paper, we build on this approach and discretise the reaction-diffusion model (\ref{PDE: Governing Reaction-Diffusion})--(\ref{EQS: Governing BC2}) in space and time to derive probabilities governing the movement, binding and release of individual drug particles. In this section, we keep the discussion focussed on the semi-permeable coating problem (second boundary condition listed in (\ref{EQS: Governing BC2})) and point out important differences when required for the fully-permeable coating problem (first boundary condition listed in (\ref{EQS: Governing BC2})). Our random walk model considers $N_{p}$ individual drug particles with the discretisation of (\ref{PDE: Governing Reaction-Diffusion})--(\ref{EQS: Governing BC2}) carried out using uniform spatial and temporal grids, defined by $r_{i} = (i-1)h$ for $i = 1,\hdots,M$ and $t_{n} = n\tau$ for $n = 1,\hdots,N_t$, where $h = R/(M-1)$ and $\tau = T/N_t$. To perform the spatial discretisation, we use the standard vertex-centred finite volume method outlined in \ref{APPENDIX: Finite volume method} with second-order central differences used to approximate derivative terms and finite volumes spanning the intervals $r\in[w_{i},e_{i}]$ for $i = 1,\hdots,M$, where $w_{i} = \max\{0,(i-3/2)h\}$ and $e_{i} = \min\{(i-1/2)h,R\}$. For the semi-permeable coating problem, this discretisation in space yields a system of ODEs,
\begin{align}
    \label{ODE: Finite Volume - Markov Interpretation}
    \frac{\diff \vec{c}}{\diff t} &= \vec{A}_\vec{m} \vec{c} - \vec{A}_{\vec{b}} \vec{c} - \vec{A}_\vec{r} \vec{c},\qquad \vec{c}(0) = \vec{c}_{0},
\end{align}
where $\vec{c} = (c_1,\hdots,c_{M})^\T$ such that $c_i \approx c(r_i,t)$, $\vec{c}_{0} = (c_{0}(r_{1}),\hdots,c_{0}(r_{M}))^\T$, and $\vec{A}_\vec{m}$, $\vec{A}_\vec{b}$, and $\vec{A}_\vec{r}$ denote, respectively, the $M\times M$ discretisation matrices governing movement (diffusion term in equation (\ref{PDE: Governing Reaction-Diffusion})), binding (reaction term in equation (\ref{PDE: Governing Reaction-Diffusion})) and release (boundary condition (\ref{EQS: Governing BC2})). As evident in \ref{APPENDIX: Finite volume method}, $\vec{A}_\vec{m}$ and $\vec{A}_{\vec{b}}$ are tridiagonal and diagonal matrices, respectively, while $\vec{A}_\vec{r}$ contains a single non-zero entry in row $M$ and column~$M$.

To study the behaviour of individual drug particles, we reformulate the problem using the relationship $c_{i} = N_{i}/(S_{p}V_{i})$, where $N_{i}$ is the number of drug particles within control volume $i$ at time $t$ (assumed located at lattice site $i$), $V_{i} = (e_{i}^{d}-w_{i}^{d})/d$ and $S_{p} = \sum_{i=1}^{M}c_{0}(r_{i})V_{i}/N_{p}$ is a scaling constant depending on the initial concentration profile $c_{0}(r)$ (\ref{EQ: Governing IC}). Applying both this relationship and a forward Euler time discretisation to equation (\ref{ODE: Finite Volume - Markov Interpretation}) yields the following scheme relating the spatial distributions of drug particles at times $t = t_{n+1}$ and $t = t_{n}$:
\begin{align}
    \label{EQ: Matrix Eqn - [Simplified] Number of Particles at t(n+1)}
    \vec{N}_{n+1}^\T &= \vec{N}_n^\T\,\vec{P}_\vec{m} - \vec{N}_n^\T\,\vec{P}_\vec{b} - \vec{N}_n^\T\, \vec{P}_\vec{r},
\end{align}
where $\vec{N} = (N_{1},\hdots,N_{M})^{\T}$, $\vec{N}_{k}$ denotes $\vec{N}$ at time $t = k\tau$, $\vec{N}_{0} = \text{round}(S_{p}(c_{0}(r_{1})V_{1},\hdots,c_{0}(r_{M})V_{M})^{\T})$, $\vec{P}_\vec{m} = \vec{V}^{-1} \left( \vec{I} + \tau \vec{A}^{\T}_\vec{m} \right) \vec{V}$, $\vec{P}_\vec{b} = \tau {\vec{A}^\T_\vec{b}}$, $\vec{P}_\vec{r} = \tau {\vec{A}^\T_\vec{r}}$ and $\vec{V} = \text{diag}\left( V_1,\hdots,V_M \right)$. As we discuss below, the entries of $\vec{P}_\vec{m}$, $\vec{P}_\vec{b}$ and $\vec{P}_\vec{r}$ can be interpreted, respectively, as probabilities governing movement, binding and release events \cite{Carr:2025_Stochastic} provided the time step $\tau$ is constrained appropriately.

\bigskip\noindent\textit{Movement probabilities}: $\vec{P}_\vec{m}$ is a tridiagonal matrix with non-zero entries: 
\begin{gather*}
    p^{m}_{{1,1}} = 1 - \frac{\tau D(e_1) e_1^{d-1}}{V_1 h}, \quad p^{m}_{{1,2}} =  \frac{\tau D(e_1) e_1^{d-1}}{V_1 h}, \\
    p^{m}_{i,i-1} = \frac{\tau D(w_i) w_i^{d-1}}{V_i h}, \quad p^{m}_{{i,i}} = 1-\frac{\tau [D(w_i) w_i^{d-1} + D(e_i) e_i^{d-1}]}{V_i h}, \quad p^{m}_{{i,i+1}} = \frac{\tau D(e_i) e_i^{d-1}}{V_i h}, \quad \text{for $i=2,...,M-1$}, \\
    p^{m}_{{M,M-1}} = \frac{\tau D(w_M) w_M^{d-1}}{V_M h}, \quad p^{m}_{{M,M}} = 1-\frac{\tau D(w_M) w_M^{d-1}}{V_M h},
\end{gather*}
where we interpret $p^{m}_{i,j}$ (entry in row $i$ and column $j$) as the probability that a drug particle located at lattice site $i$ at time $t=t_n$ moves to lattice site $j$ (remains at lattice site $j$ if $i=j$) at time $t=t_{n+1}=t_n+\tau$ (given that it did not undergo a binding or release event at time $t=t_n$). Studying the above expressions, we note that each row of $\vec{P}_{\vec{m}}$ sums to 1, which is an expected result as drug particles must undergo a movement event if they did not undergo a binding or release event. For the fully-permeable coating problem, we note further that the movement probabilities apply only at lattice sites $i = 1,\hdots,M-1$, since all drug particles located at lattice site $N$ at time $t = t_{n}$ are immediately released from the device with probability one.

\bigskip\noindent\textit{Binding probabilities}: $\vec{P}_\vec{b}$ is a diagonal matrix with non-zero entries:
\begin{align*}
p^{b}_{{i,i}}=\tau k(r_i), \quad \text{for $i=1,...,M$},
\end{align*}
where we interpret $p^{b}_{i,i}$ (entry in row $i$ and column $i$) as the probability that a drug particle located at lattice site $i$ at time $t=t_n$ binds to the carrier material and remains at lattice site $i$ thereafter (given that it did not undergo a release event at time $t=t_n$). For the fully-permeable coating problem, we note that the binding probabilities apply only at lattice sites $i = 1,\hdots,M-1$, since all drug particles located at lattice site $M$ at time $t=t_{n}$ are immediately released from the device with probability one.

\bigskip\noindent\textit{Release probabilities}: $\vec{P}_\vec{r}$ is a matrix with a single non-zero entry: 
\begin{align*}
    p^{r}_{M,M} = \frac{\tau P R^{d-1}}{V_M},
\end{align*}
where we interpret $p^{r}_{M,M}$ (entry in row $M$ and column $M$) as the probability that a drug particle located at lattice site $M$ at time $t = t_{n}$ is released from the device (and no longer undergoes any movement or binding events). Note that $p^{r}_{M,M}$ increases for increasing values of $P$, which agrees intuitively with larger values of the mass transfer coefficient increasing the rate at which drug is released from the device. For the fully-permeable coating problem, we simply take $p^{r}_{M,M} = 1$, since all drug particles located at lattice site $M$ at time $t=t_{n}$ are immediately released from the device with probability one.

\subsection{Time step constraints}
For a valid random walk model we must ensure that all entries of $\vec{P}_\vec{m}$, $\vec{P}_\vec{b}$, and $\vec{P}_\vec{r}$ are in $[0,1]$. Studying the probabilities given in the previous section, this yields a set of conditions on the time step, $\tau$, depending on the coating permeability that must be satisfied simultaneously:

\bigskip\noindent\textit{Fully-permeable coating}:
\begin{align*}
\tau &\leq\min\left\{\frac{V_1 h}{D(e_1) e_1^{d-1}},\min\left\{\frac{V_i h}{D(w_i) w_i^{d-1} + D(e_i) e_i^{d-1}}, \ i=2,\hdots,M-1\right\}\right\},\\
\tau &\leq\min\left\{\frac{1}{k(r_i)}, \ i=1,\hdots,M-1\right\},\\[1ex]
\intertext{\textit{Semi-permeable coating}:}
\tau &\leq \min\left\{\frac{V_1 h}{D(e_1) e_1^{d-1}},\min\left\{\frac{V_i h}{D(w_i) w_i^{d-1} + D(e_i) e_i^{d-1}}, \ i=2,\hdots,M-1\right\},\frac{V_M h}{D(w_M) w_M^{d-1}}\right\},\\
\tau &\leq \min\left\{\frac{1}{k(r_i)}, \ i=1,\hdots,M\right\},\\
\tau &\leq \frac{V_M}{PR^{d-1}}.
\end{align*}
Note that the conditions for the fully-permeable coating are less restrictive than those for the semi-permeable coating due to the absence of any conditions arising from lattice node $M$, where drug particles are immediately released from the device.

\subsection{Random walk algorithm}
Using the stochastic-discrete approach, the fraction of drug released over time can be calculated by simply tracking the cumulative number of drug particles released over time. Formally, we have
\begin{align*}
F(t_{n}) = \frac{N_{r}(t_{n})}{N_{p}},\qquad \text{for $n = 1,\hdots,N_t$},
\end{align*}
where $N_{r}(t_{n})$ is the number of drug particles released between $t= 0$ and $t = t_{n}$ and $N_{p}$ is the total number of drug particles initially loaded into the device. The total fraction of drug released is then given by $F_{\infty} = F(t_{N_{t}}) = F(T)$, where we assume $T$ (or equivalently $N_{t}$) is large enough to ensure there are no longer any active drug particles remaining in the device (all particles have undergone either a binding event or a release event). Algorithm \ref{ALG: stochastic_model} summarises the movement, binding and release of individual drug particles and the stochastic-discrete approach for calculating the fraction of drug released.

\bigskip\noindent
\algbox{
\begin{algorithm}[Random Walk Model]\mbox{}\\
\label{ALG: stochastic_model}
\emph{
\begin{tabular}{@{}l}
$C_{i,0} = c_0(r_i)$ for all $i=1,...,M$ \comment{initial concentration at lattice site $i$} \\
$S = \sum_{i=1}^{M} C_{i,0}V_i/N_p$ \comment{scaling constant} \\
$N_{i,0} = \text{round}(C_{i,0}V_i/S)$ for all $i=1,...,M$ \comment{initial number of particles at lattice site $i$} \\
$N_p = \sum_{i=1}^{M} N_{i,0}$ \comment{adjusted total number of particles} \\
$k=0$ \comment{initialise particle index} \\
\textbf{for} $i=1,...,M$ \comment{loop over lattice sites} \\
\qquad $X_{p,0} = i$ for all $p=k+1,...,k+N_{i,0}$ \comment{initial lattice site of particle $p$} \\
\qquad $k = k+N_{i,0}$ \comment{update particle index} \\
\textbf{end} \\
$N_r(t_0) = 0$ \comment{initial number of released particles at $t=t_0=0$} \\
$A_p = \textbf{true}$ for all $p = 1,...,N_p$ \comment{set active status as initially true for all particles} \\
$P_{i,0}=0$ and $P_{i,j} = \sum_{z=1}^j p^m_{i,z}$ for all $i=1,...,M$ and $j=1,...,M$ \comment{cumulative movement probabilities} \\
\textbf{for} $n=1,...,N_t$ \comment{loop over time steps} \\
\qquad $N_r(t_n) = N_r(t_{n-1})$ \comment{number of particles released at $t = t_{n-1}$}\\
\qquad \textbf{for} $p = 1,...,N_p$ \comment{loop over number of particles} \\
\qquad \qquad \textbf{if} $A_p$ \textbf{true} and $X_{p,n-1}=M$ \comment{check if particle is currently active and at end lattice site} \\
\qquad \qquad \qquad \textbf{if} fully-permeable coating \\
\qquad \qquad \qquad \qquad $N_r(t_n) = N_r(t_{n})+1$ and $A_p = \textbf{false}$ \comment{release particle} \\
\qquad \qquad \qquad \textbf{else if} semi-permeable coating \\
\qquad \qquad \qquad \qquad Sample $q \sim \text{Uniform}(0,1)$ \comment{uniform random number in $[0,1]$} \\
\qquad \qquad \qquad \qquad If $q \in (0,p^r_{M,M})$ then $N_r(t_n) = N_r(t_{n})+1$ and $A_p = \textbf{false}$ \comment{release particle}\\
\qquad \qquad \qquad \textbf{end} \\
\qquad \qquad \textbf{end} \\
\qquad \qquad \textbf{if} $A_p$ \textbf{true} \comment{check if particle is still currently active} \\
\qquad \qquad \qquad Sample $q \sim \text{Uniform}(0,1)$ \comment{uniform random number in $[0,1]$} \\
\qquad \qquad \qquad If $q \in (0,p^b_{i,i})$ then $A_p = \textbf{false}$ \comment{bind particle} \\
\qquad \qquad \textbf{end} \\
\qquad \qquad \textbf{if} $A_p$ \textbf{true} \comment{check if particle is still currently active} \\
\qquad \qquad \qquad Sample $q \sim \text{Uniform}(0,1)$ \comment{uniform random number in $[0,1]$} \\
\qquad \qquad \qquad Find $j$ such that $q \in (P_{i,j-1},P_{i,j})$, then $X_{p,n}=j$ \comment{move particle} \\
\qquad \qquad \textbf{end} \\
\qquad \textbf{end} \\
\qquad $F(t_n) = N_r(t_n)/N_p$ \comment{fraction of drug released at $t = t_{n}$} \\
\textbf{end} \\
$F_{\infty} = F(t_n)$ \comment{total fraction of drug released}
\end{tabular}
}
\end{algorithm}}

\section{Results}
\label{SEC: Results}

\subsection{Diffusivity and reaction-rate functions}

Applying the deterministic-continuum (section \ref{SEC: Analytical Approach}) and stochastic-discrete (section \ref{SEC: Stochastic Approach}) approaches requires choosing appropriate diffusivity and reaction-rate functions, $D(r)$ and $k(r)$, to characterise the heterogeneity of the FGM. In this work, we use standard smooth approximations to the step-wise functions
\begin{align}
    \label{EQ: Step-Wise Functions - D}
    D(r) &= \begin{cases} D_{\rm{max}}, & 0<r<\sigma, \\ D_{\rm{min}}, & \sigma<r<R, \end{cases}\\
    \label{EQ: Step-Wise Functions - k}
    k(r) &= \begin{cases} k_{\rm{min}}, & 0<r<\sigma, \\ k_{\rm{max}}, & \sigma<r<R, \end{cases}
\end{align}
where $D_{\rm{max}}>D_{\rm{min}}$ and $k_{\rm{max}}>k_{\rm{min}}$. Following previous work \cite{Carr:2024_FunctionalRelease}, we choose
\begin{align}
    \label{EQ: Smooth Step-Wise Functions - D}
    D(r) &= D_{\rm{max}} + (D_{\rm{min}} - D_{\rm{max}}) \left( \frac{1}{2} + \frac{1}{\pi} \arctan\left( \frac{\alpha(r-\sigma)}{R} \right) \right), \\
    \label{EQ: Smooth Step-Wise Functions - k}
    k(r) &= k_{\rm{min}} + (k_{\rm{max}} - k_{\rm{min}}) \left( \frac{1}{2} + \frac{1}{\pi} \arctan\left( \frac{\alpha(r-\sigma)}{R} \right) \right),
\end{align}
where $\alpha$ and $\sigma$ are parameters controlling the steepness and location of the transition between the stepwise values, respectively. Note that $D(r)$ and $k(r)$ are decreasing and increasing functions of $r$, respectively, and can be considered as corresponding to an FGM exhibiting a decreasing porosity as $r$ increases \cite{Carr:2024_FunctionalRelease}.

\begin{figure}[t]
    \centering
    Diffusivity\\[0.2cm]
    \begin{subfigure}{0.3\textwidth}
        \centering
        \includegraphics[width=\textwidth]{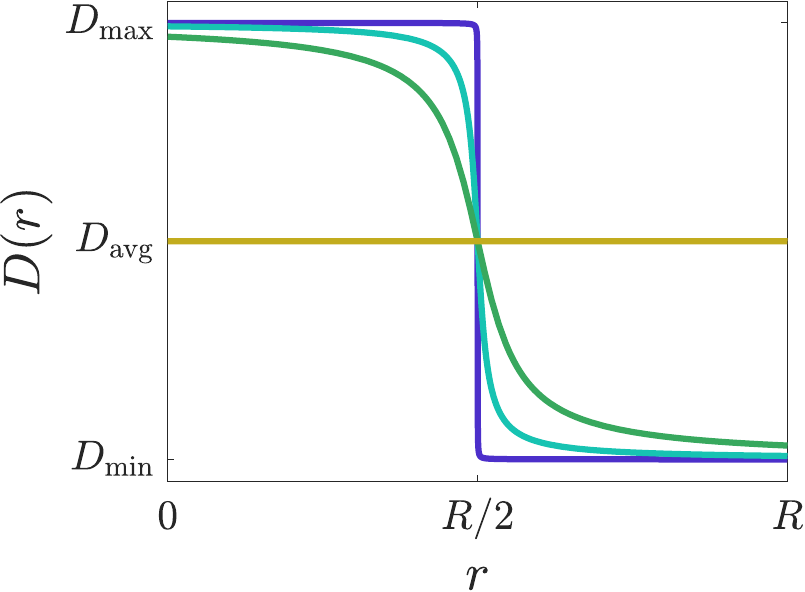}
    \end{subfigure}
    \hspace{10px}
    \begin{subfigure}{0.3\textwidth}
        \centering
        \includegraphics[width=\textwidth]{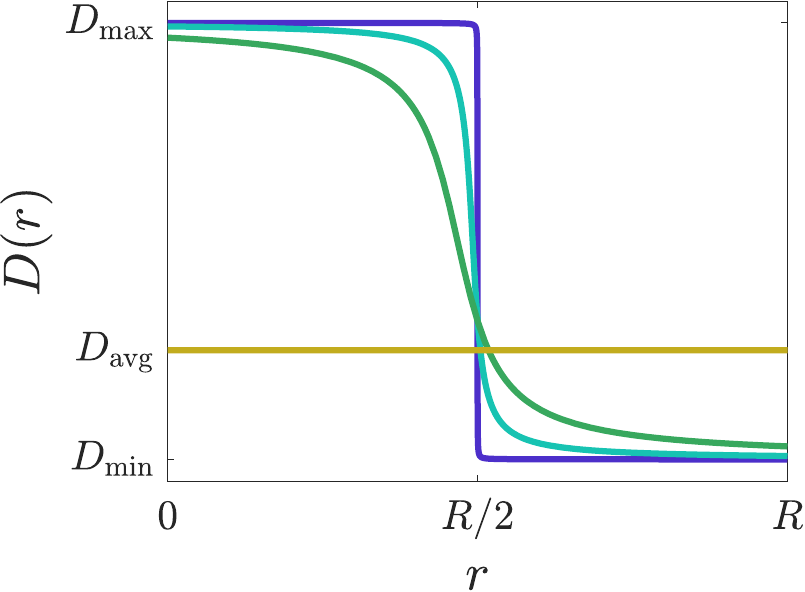}
    \end{subfigure}
    \hspace{10px}
    \begin{subfigure}{0.3\textwidth}
        \centering
        \includegraphics[width=\textwidth]{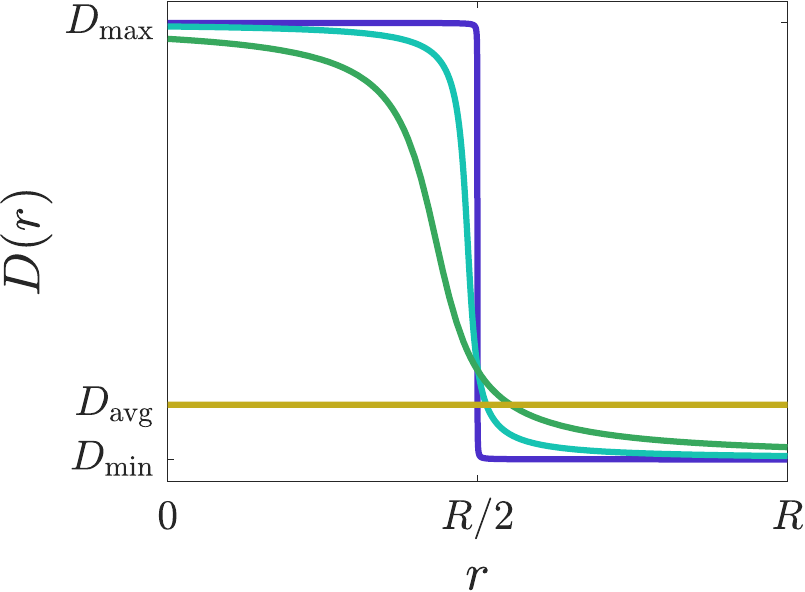}\\[0.2cm]
    \end{subfigure}
    Reaction-rate\\[0.2cm]
    \begin{subfigure}{0.3\textwidth}
        \centering
        \includegraphics[width=\textwidth]{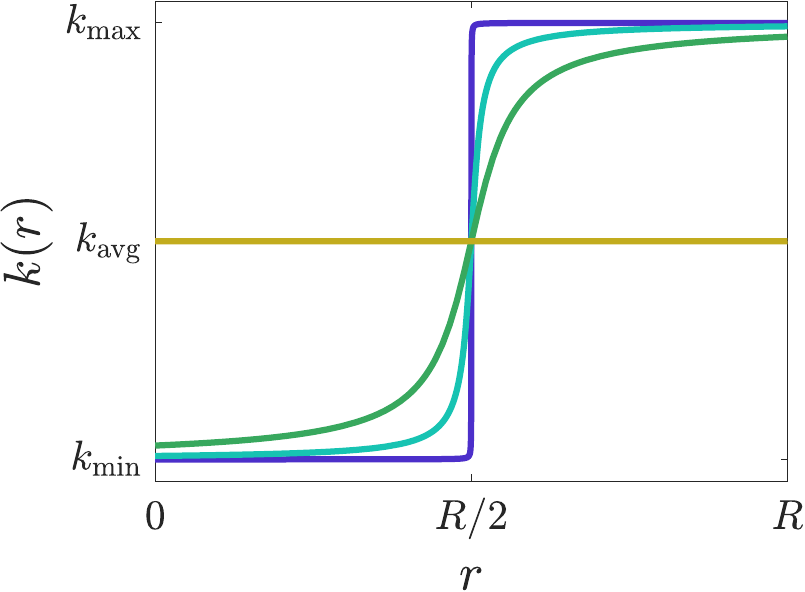}
        \caption{\normalsize Slab [$d=1$]}
    \end{subfigure}
    \hspace{10px}
    \begin{subfigure}{0.3\textwidth}
        \centering
        \includegraphics[width=\textwidth]{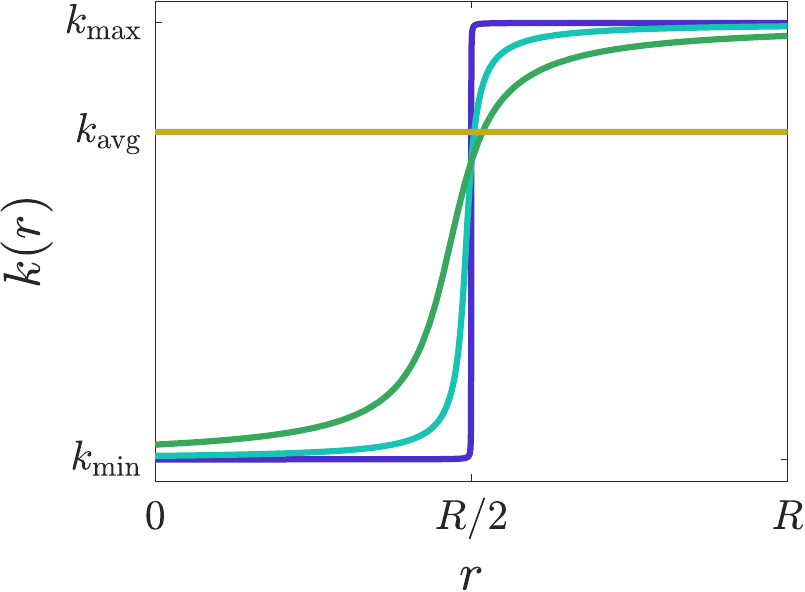}
        \caption{\normalsize Cylinder [$d=2$]}
    \end{subfigure}
    \hspace{10px}
    \begin{subfigure}{0.3\textwidth}
        \centering
        \includegraphics[width=\textwidth]{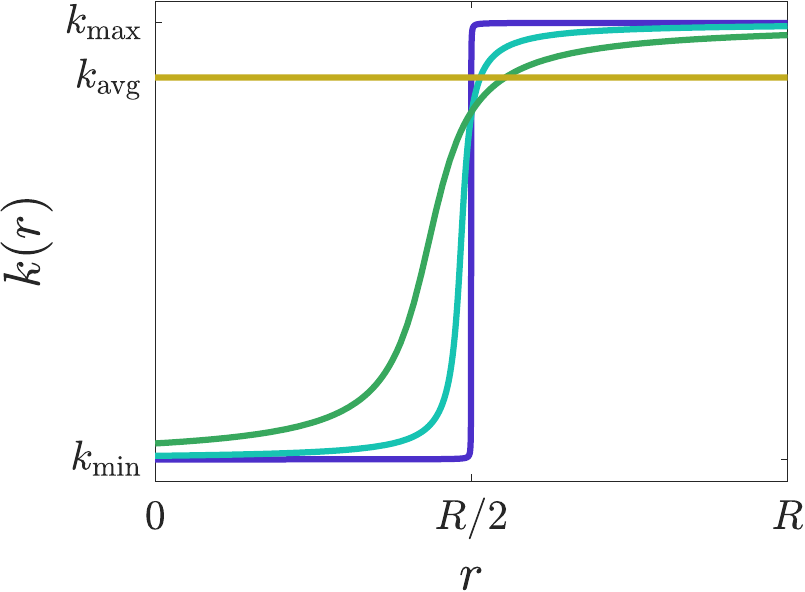}
        \caption{\normalsize Sphere [$d=3$]}
    \end{subfigure}
    \vspace*{0.3cm}
    
    \definecolor{alphaYellow}{RGB}{194, 172, 30}
    \definecolor{alphaGreen}{RGB}{56, 168, 94}
    \definecolor{alphaCyan}{RGB}{23, 195, 178}
    \definecolor{alphaBlue}{RGB}{76, 47, 202}
    \begin{minipage}{0.9\textwidth}
        \centering
        \begin{tabular}{c c c c c}
            \hhline{-----}
            \hspace{-10px} & $\alpha$ & \rule{0pt}{2.2ex} $\hat{\sigma}$ [$d=1$] & $\hat{\sigma}$ [$d=2$] & $\hat{\sigma}$ [$d=3$] \\
            \hhline{-----}
            \textcolor{alphaYellow}{\rule{1.5ex}{1.5ex}} \hspace{-10px} & \rule{0pt}{2.2ex} 0.0001 & $0.5$ & $-9999$ & $-24140$ \\
            \textcolor{alphaGreen}{\rule{1.5ex}{1.5ex}} \hspace{-10px} & \rule{0pt}{2.2ex} 20 & $0.5$ & $0.4676$ & $0.4330$ \\
            \textcolor{alphaCyan}{\rule{1.5ex}{1.5ex}} \hspace{-10px} & \rule{0pt}{2.2ex} 80 & $0.5$ & $0.4920$ & $0.4839$ \\
            \textcolor{alphaBlue}{\rule{1.5ex}{1.5ex}} \hspace{-10px} & \rule{0pt}{2.2ex} 10000 & 0.5 & 0.4998 & 0.4997 \\
            \hhline{-----}
        \end{tabular}
    \end{minipage}
    \caption{\textbf{Functionally-graded diffusivity and reaction-rate functions.} Diffusivity $D(r)$ (top row) and reaction-rate $k(r)$ (bottom row) profiles for each device-geometry (slab, cylinder, sphere). For each value of $\alpha$ and $d$, the table provides the dimensionless value of $\sigma$ satisfying equation (\ref{EQ: Nonlinear equation - solving sigma}), defined by $\hat{\sigma} = \sigma/R$, rounded to four significant digits. Parameters: $[D_{\rm{min}}, D_{\rm{max}}] = [10^{-13}, 10^{-11}]\,\text{cm}^{2}\,\text{s}^{-1}$ and $[k_{\rm{min}}, k_{\rm{max}}] = [8\cdot10^{-5}, 10^{-4}]\,\text{s}^{-1}$.}
    \label{FIG: Parameter Profiles - D and k}
\end{figure}

For a specified value of $\alpha$, we follow \cite{Carr:2024_FunctionalRelease} and determine a corresponding value of $\sigma$ to ensure the average diffusivity across the slab ($d=1$), cylinder ($d=2$) and sphere ($d=3$) geometries remains fixed. In particular, for each $d$, we enforce that
\begin{align}
    \label{EQ: Diffusivity Average - Integral Form}
    \frac{d}{R^d} \int_0^R r^d D(r) \, \diff r = D_{\rm{avg}},
\end{align}
where $D_{\rm{avg}}$ is chosen as the average diffusivity of the step-wise function (\ref{EQ: Step-Wise Functions - D}) over $[0,R]$ when $\sigma = R/2$:
\begin{align}
    \label{EQ: Diffusivity Average - Simplified Form}
    D_{\rm{avg}} = \frac{d}{R^d} \left( \int_0^{R/2} r^d D_{\rm{max}} \, \diff r + \int_{R/2}^{R} r^d D_{\rm{min}} \, \diff r \right) = \frac{1}{2^d} D_{\rm{max}} + \left( 1 - \frac{1}{2^d} \right) D_{\rm{min}}.
\end{align}
For a specified value of $\alpha$, we therefore determine a corresponding value of $\sigma$ by solving the following nonlinear equation (for further details on the numerical procedure employed, the interested reader is referred to our MATLAB code available on GitHub; see \nameref{sec:data_availability}):
\begin{align}
    \label{EQ: Nonlinear equation - solving sigma}
    \frac{d}{R^d} \int_0^R r^d \left( D_{\rm{max}} + (D_{\rm{min}} - D_{\rm{max}}) \left( \frac{1}{2} + \frac{1}{\pi} \arctan\left( \frac{\alpha(r-\sigma)}{R} \right) \right) \right) \, \diff r = D_{\rm{avg}}.
\end{align}
In the case of $k(r)$, the same values for $\alpha$ and $\sigma$ are used, which equivalently ensures that the average reaction rate is maintained across the slab ($d=1$), cylinder ($d=2$) and sphere ($d=3$) geometries, that is,
\begin{align}
    \label{EQ: Reaction-rate Average - Integral Form}
    \frac{d}{R^d} \int_0^R r^d k(r) \, \diff r = k_{\rm{avg}},
\end{align}
where $k_{\rm{avg}}$ is the average reaction-rate of the stepwise function (\ref{EQ: Step-Wise Functions - k}) over $[0,R]$ when $\sigma = R/2$:
\begin{align}
    \label{EQ: Reaction-rate Average - Simplified Form}
    k_{\rm{avg}} = \frac{1}{2^d} k_{\rm{min}} + \left( 1 - \frac{1}{2^d} \right) k_{\rm{max}}.
\end{align}
For each of the slab ($d=1$), cylinder ($d=2$) and sphere ($d=3$) geometries, Figure~\ref{FIG: Parameter Profiles - D and k} plots the diffusivity and reaction-rate functions, $D(r)$ and $k(r)$, obtained using the above procedure for four different choices of $\alpha$ and set parameter values for $[D_{\text{min}},D_{\text{max}}]$ and $[k_{\text{min}},k_{\text{max}}]$ (Figure \ref{FIG: Parameter Profiles - D and k}) \cite{Carr:2024_FunctionalRelease}. Note that the smallest and largest values of $\alpha$ accurately approximate the two limiting cases of a homogeneous system with constant $D(r)$ and $k(r)$ ($\alpha = 0.0001$) and a composite system with stepwise $D(r)$ and $k(r)$ ($\alpha = 10000$). 

\subsection{Verification of deterministic-continuum and stochastic-discrete approaches}
\label{SEC: Verification}
\begin{figure}[p]
    \centering
    \def\figw{0.39\textwidth}
    \begin{subfigure}{\figw}
        \centering
        \includegraphics[width=\textwidth]{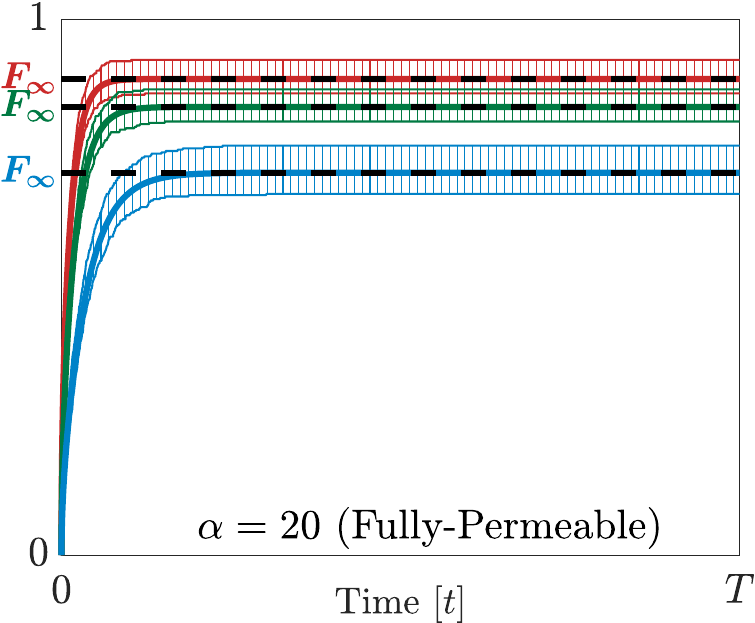}
    \end{subfigure}
    \hspace{10px}
    \begin{subfigure}{\figw}
        \centering
        \includegraphics[width=\textwidth]{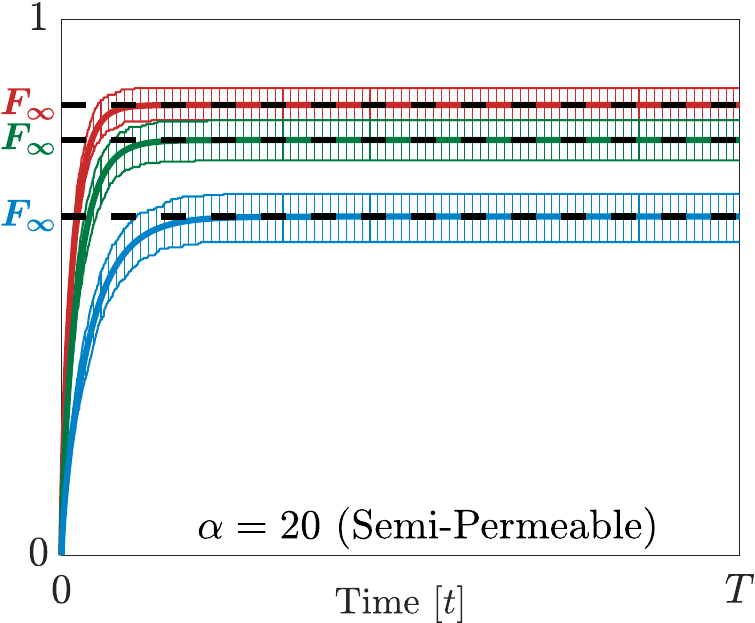}
    \end{subfigure}
    \hspace{10px}
    \begin{subfigure}{\figw}
        \centering
        \includegraphics[width=\textwidth]{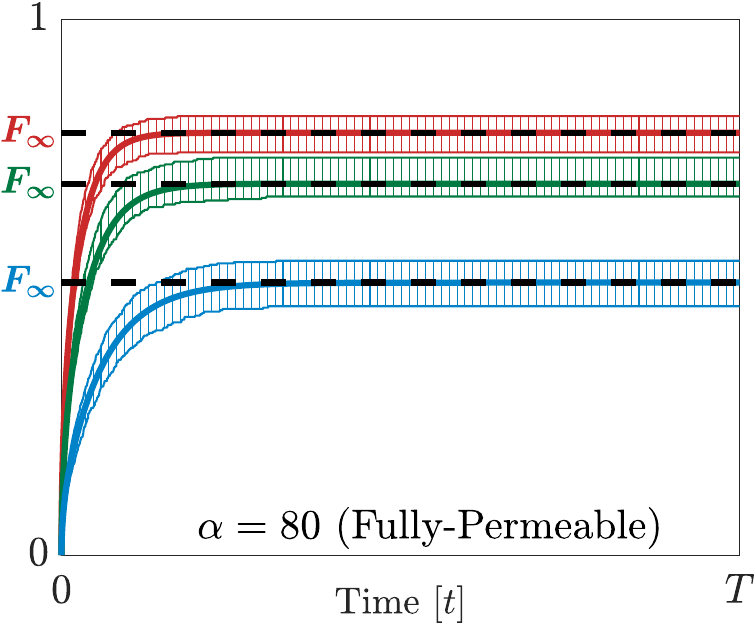}
    \end{subfigure}
    \hspace{10px}
    \begin{subfigure}{\figw}
        \centering
        \includegraphics[width=\textwidth]{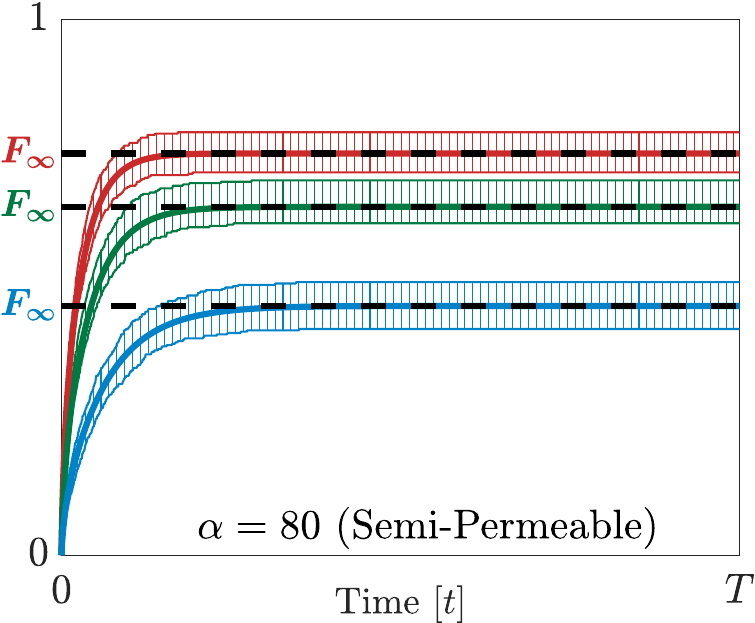}
    \end{subfigure}
    \hspace{10px}
    \begin{subfigure}{\figw}
        \centering
        \includegraphics[width=\textwidth]{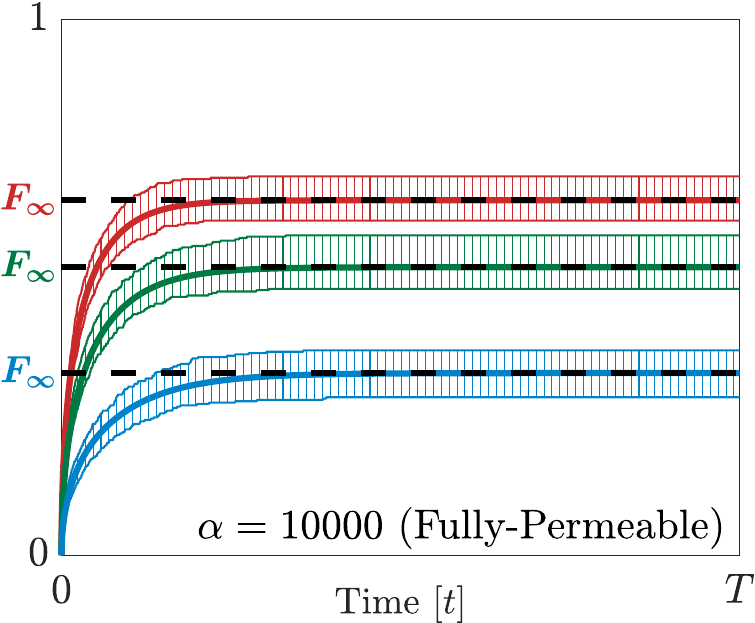}
    \end{subfigure}
    \hspace{10px}
    \begin{subfigure}{\figw}
        \centering
        \includegraphics[width=\textwidth]{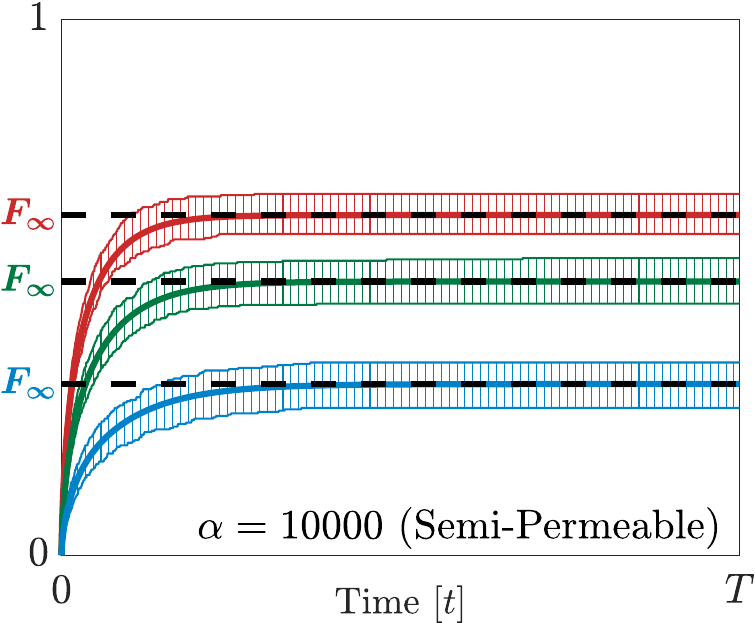}
    \end{subfigure}
    \vspace*{0.4cm}
    
    \definecolor{dRed}{RGB}{203, 42, 42}
    \definecolor{dGreen}{RGB}{0, 122, 67}
    \definecolor{dBlue}{RGB}{0, 130, 200}
    \begin{minipage}{1\textwidth}
        \centering
        \begin{tabular}{c l c c c c c c}
            \hhline{--------}
            \multicolumn{2}{@{}l}{} & \multicolumn{3}{l}{Fully-Permeable} & \multicolumn{3}{l}{Semi-Permeable} \\
            \hline
            $\alpha$ & Method & \textcolor{dBlue}{\rule{1.5ex}{1.5ex}} $d=1$ & \textcolor{dGreen}{\rule{1.5ex}{1.5ex}} $d=2$ & \textcolor{dRed}{\rule{1.5ex}{1.5ex}} $d=3$ & \textcolor{dBlue}{\rule{1.5ex}{1.5ex}} $d=1$ & \textcolor{dGreen}{\rule{1.5ex}{1.5ex}} $d=2$ & \textcolor{dRed}{\rule{1.5ex}{1.5ex}} $d=3$ \\
            \hline
            0.0001 & \makecell{Analytical\\ Numerical} & \makecell{0.9445\\0.9445} & \makecell{0.9566\\0.9566} & \makecell{0.9546\\0.9546} & \makecell{0.8073\\0.8073} & \makecell{0.8769\\0.8769} & \makecell{0.8988\\0.8988} \\
            \hline
            20 & \makecell{Analytical\\ Numerical} & \makecell{0.7145\\0.7145} & \makecell{0.8372\\0.8372} & \makecell{0.8894\\0.8894} & \makecell{0.6325\\0.6325} & \makecell{0.7753\\0.7753} & \makecell{0.8407\\0.8407} \\
            \hline
            80 & \makecell{Analytical\\ Numerical} & \makecell{0.5091\\0.5091} & \makecell{0.6940\\0.6940} & \makecell{0.7891\\0.7891} & \makecell{0.4656\\0.4656} & \makecell{0.6507\\0.6507} & \makecell{0.7505\\0.7505} \\
            \hline
            10000 & \makecell{Analytical\\ Numerical} & \makecell{0.3402\\0.3402} & \makecell{0.5378\\0.5378} & \makecell{0.6629\\0.6629} & \makecell{0.3196\\0.3196} & \makecell{0.5112\\0.5112} & \makecell{0.6352\\0.6352} \\
            \hline
        \end{tabular}
    \end{minipage}
    \caption{\textbf{Verification of deterministic-continuum and stochastic-discrete approaches.} Plots depict the analytical values of $F_{\infty}$ computed using equations (\ref{EQ: Finf Final Result Fully - d=1})--(\ref{EQ: Finf Final Result Semi - d=3}) (dashed line) and stochastic-discrete profiles of $F(t)$ computed using Algorithm \ref{ALG: stochastic_model} (enveloping regions), benchmarked against numerical approximations to $F(t)$ as per \ref{APPENDIX: Numerical Computation} (solid line). For the stochastic-discrete approach, the enveloping regions extend between the 10\% and 90\% quantiles across $100$ trials of Algorithm \ref{ALG: stochastic_model}. Table compares the analytical values of $F_{\infty}$ computed using equations (\ref{EQ: Finf Final Result Fully - d=1})--(\ref{EQ: Finf Final Result Semi - d=3}) to benchmark numerical values of $F_{\infty}$ computed as per \ref{APPENDIX: Numerical Computation}. Results are given for each combination of device geometry (\textcolor[RGB]{0,130,200}{\rule{0.7em}{0.7em}} $d=1$ (slab); \textcolor[RGB]{0,122,67}{\rule{0.7em}{0.7em}} $d=2$ (cylinder); \textcolor[RGB]{203,42,42}{\rule{0.7em}{0.7em}} $d=3$ (sphere)), coating permeability (fully-permeable, semi-permeable), and $\alpha$ value (prescribing $D(r)$ and $k(r)$ as in Figure \ref{FIG: Parameter Profiles - D and k}). Note colour coding for $d=1,2,3$ applies across all plots and the table. The profiles for $\alpha=0.0001$ show equivalent levels of agreement but are not shown since the stochastic-discrete enveloping regions greatly overlap when visualised. Parameters:  $N_{s} = 400$ (analytical); $M = 51$, $N_{t} = 1.5\cdot 10^{6}$, $N_{p} = 200$ (stochastic); $M=5001$, $N_{t}=10^{4}$ (numerical); $c_0(r)=0.4\,\text{mol}\,\text{cm}^{-3}$, $R = 10^{-4}\,\text{cm}$, $P=5\cdot10^{-8}\,\text{cm}\,\text{s}^{-1}$, $T=10^5\,\text{s}$.}
    \label{FIG: Release Profiles}
\end{figure}

We now verify our deterministic-continuum (section \ref{SEC: Analytical Approach}) and stochastic-discrete (section \ref{SEC: Stochastic Approach}) approaches for calculating the fraction of drug released. In total, we consider 24 different test cases, one for each combination of device geometry (slab, cylinder, sphere), coating permeability (fully-permeable, semi-permeable) and $\alpha$ value ($\alpha = 0.0001$, $20$, $80$, $10000$; see Figure \ref{FIG: Parameter Profiles - D and k}). For each test case, we compare the value of $F_{\infty}$ (total fraction of drug released) obtained from the deterministic-continuum approach via the analytical formulas (\ref{EQ: Finf Final Result Fully - d=1})--(\ref{EQ: Finf Final Result Semi - d=3}) to profiles of $F(t)$ (fraction of drug released over time) obtained from the stochastic-discrete approach via repeated simulations of the random walk algorithm (Algorithm \ref{ALG: stochastic_model}). Both approaches are also benchmarked against numerical approximations of $F_{\infty}$ and $F(t)$, calculated by combining a numerical solution of the reaction-diffusion model (\ref{PDE: Governing Reaction-Diffusion})--(\ref{EQS: Governing BC2}) with the following analytical expressions for $F_{\infty}$ and $F(t)$ (see \ref{APPENDIX: Numerical Computation} for further details):
\begin{align}
    \label{EQ: Finf and Ft - c}
    F_{\infty} = \lim_{t\rightarrow\infty} F(t),\quad F(t) = \frac{R^{d-1} \int_0^{t} -D(R) \frac{\partial c}{\partial r}(R,s) \, \diff s}{\int_0^R r^{d-1} c_0(r) \, \diff r}.
\end{align}
Results presented in Figure \ref{FIG: Release Profiles} show good agreement between all methods for calculating the fraction of drug released. In particular, for each of the 24 test cases: the deterministic-continuum (analytical) value of $F_{\infty}$ falls within the limiting behaviour of the stochastic-discrete profile of $F(t)$; the deterministic-continuum (analytical) value of $F_{\infty}$ agrees with the numerical estimate of $F_{\infty}$ to four significant figures; and the stochastic-discrete profile of $F(t)$ envelopes the numerical profile of $F(t)$. In summary, these results provide strong evidence to support the accuracy of both the deterministic-continuum and stochastic-discrete approaches developed in this paper.

In addition to verifying our mathematical working, several physical observations can also be drawn from the results presented in Figure~\ref{FIG: Release Profiles}. First, we see that $F_{\infty}$ increases when $d$ increases, which is explained by the fact that drug particles are more likely to move outwards towards the release surface ($r = R$) than inwards for cylinder and sphere geometries ($d = 2,3$) and the fact that this biased movement towards the release surface  is more pronounced when $d$ increases \cite{Carr:2020_Diffusion}. Second, we see that $F_{\infty}$ is larger for fully-permeable coatings than for semi-permeable coatings. This is easily explained by the fact that semi-permeable coatings reject the release of some drug particles at the release surface ($r=R$) prolonging their time within the device and therefore increasing the likelihood that they undergo a binding event and never be released. Finally, we see that $F_{\infty}$ decreases when $\alpha$ increases. This is explained by the fact that drug particles are both more likely to undergo a binding event and less likely to undergo a movement event in $(R/2,R)$ (further increasing the likelihood they undergo a binding event at some stage) when $\alpha$ is large, which results in less drug particles being released.
  
\subsection{Reproduction of previous results}
In previous work, Carr \cite{Carr:2024_FractionRelease} developed analytical expressions for $F_{\infty}$ for two special cases of the reaction-diffusion model (\ref{PDE: Governing Reaction-Diffusion})--(\ref{EQS: Governing BC2}) corresponding to a monolithic and core-shell system, respectively. The monolithic system considers constant values for the diffusivity $D(r)$, reaction-rate $k(r)$ and initial concentration $c_{0}(r)$, whereas the core-shell system considers step-wise functions (with no binding in the core):

\bigskip\noindent\textit{Monolithic system}:
\begin{gather}
\label{EQ: Monolithic}
D(r) = D, \quad k(r) = k,\quad c_{0}(r) = c_{0},\quad 0 < r < R,
\intertext{\textit{Core-shell system}:}
\label{EQ: Core-shell}
D(r) = \begin{cases} D_{\rm{c}}, & 0 < r < R_{\rm{c}},\\ D_{\rm{s}}, & R_{\rm{c}} < r < R,\end{cases}\quad
k(r) = \begin{cases} 0, & 0 < r < R_{\rm{c}},\\ k_{\rm{s}}, & R_{\rm{c}} < r < R,\end{cases}\quad c_{0}(r) = \begin{cases} c_{0}, & 0 < r < R_{\rm{c}},\\ 0, & R_{\rm{c}} < r < R.\end{cases}
\end{gather}
We now show how our formulas for $F_{\infty}$ can be applied to replicate both the monolithic and core-shell cases. For the monolithic case, we simply choose $D(r)$, $k(r)$ and $c_{0}(r)$ to be constant as in equation (\ref{EQ: Monolithic}). For the core-shell case, however, we cannot work with discontinuous step-wise $D(r)$, so we use smooth approximations to $D(r)$ and $k(r)$ as in (\ref{EQ: Smooth Step-Wise Functions - D})--(\ref{EQ: Smooth Step-Wise Functions - k}) with $D_{\rm{max}} = D_{\rm{c}}$, $D_{\rm{min}} = D_{\rm{s}}$, $k_{\rm{min}} = 0$ and $k_{\rm{max}} = k_{\rm{s}}$.

Tables \ref{TAB: Monolithic Results Comparison} and \ref{TAB: Coreshell Results Comparison} compare values of $F_{\infty}$ obtained using the analytical expressions developed in the current paper (equations (\ref{EQ: Finf Final Result Fully - d=1})--(\ref{EQ: Finf Final Result Semi - d=3})) to those obtained using the analytical expressions of Carr \cite{Carr:2024_FractionRelease} for both the monolithic and core-shell systems. For the monolithic system, we see that the reported values of $F_{\infty}$ agree to the four decimal places reported for each combination of device geometry ($d=1,2,3$ for slab, cylinder, sphere) and coating permeability (fully-permeable, semi-permeable). For the core-shell system, the values of $F_{\infty}$ computed using our formulas approach those of Carr \cite{Carr:2024_FractionRelease} as $\alpha$ increases (i.e. as $D(r)$ and $k(r)$ become closer to step-wise functions). In summary, these results provide further evidence to support the analytical expressions for $F_{\infty}$ developed.

\begin{table}[h]
    \centering
    \begin{tabular}{c c c c c c c}
        \hhline{~------}
        & \multicolumn{3}{c}{Fully-Permeable} & \multicolumn{3}{c}{Semi-Permeable} \\
        \hline
        Formula Set & \rule{0pt}{2.2ex} $d=1$ & $d=2$ & $d=3$ & $d=1$ & $d=2$ & $d=3$ \\
        \hline
        Current paper & \rule{0pt}{2.2ex} 0.8611 & 0.9423 & 0.9682 & 0.7928 & 0.8999 & 0.9379 \\
        Carr (2024) & \rule{0pt}{2.2ex} 0.8611 & 0.9423 & 0.9682 & 0.7928 & 0.8999 & 0.9379 \\
        \hline
    \end{tabular}
    \caption{\textbf{Comparison to Carr's (2024) monolithic formulas.} Calculated values of $F_{\infty}$ obtained using the analytical expressions developed in the current paper (equations (\ref{EQ: Finf Final Result Fully - d=1})--(\ref{EQ: Finf Final Result Semi - d=3})) to those obtained using the monolithic analytical expressions developed previously by Carr (2024) (equations (10)--(15) in \cite{Carr:2024_FractionRelease}). Results are given for each combination of device geometry ($d=1,2,3$ for slab, cylinder, sphere) and coating permeability (fully-permeable, semi-permeable). Parameter values: $D(r)=10^{-12}\,\text{cm}^2\,\text{s}^{-1}$, $k(r)=5\cdot 10^{-5}\,\text{s}^{-1}$, $c_{0}(r) = 0.4\,\text{mol}\,\text{cm}^{-3}$, $R = 10^{-4}\,\text{cm}$, $P = 5\cdot 10^{-8}\,\text{cm}\,\text{s}^{-1}$, $N_{s} = 400$ [Current paper]; $D = 10^{-12}\,\text{cm}^2\,\text{s}^{-1}$, $k = 5\cdot 10^{-5}\,\text{s}^{-1}$, $R = 10^{-4}\,\text{cm}$, $P = 5\cdot 10^{-8}\,\text{cm}\,\text{s}^{-1}$ [Carr (2024)].}
    \label{TAB: Monolithic Results Comparison}
\end{table}
\begin{table}[t]
    \centering
    \begin{tabular}{c c c c c c c c}
        \hhline{~~------}
        & \rule{0pt}{2.2ex} & \multicolumn{3}{c}{Fully-Permeable} & \multicolumn{3}{c}{Semi-Permeable} \\
        \hline
        Formula Set & \rule{0pt}{2.2ex} $\alpha$ & $d=1$ & $d=2$ & $d=3$ & $d=1$ & $d=2$ & $d=3$ \\
        \hline
        Current paper & \rule{0pt}{2.2ex} $10^2$ & 0.5959 & 0.6466 & 0.6885 & 0.5566 & 0.6115 & 0.6573 \\
         & \rule{0pt}{2.2ex} $10^3$ & 0.4283 & 0.4841 & 0.5345 & 0.4041 & 0.4609 & 0.5125 \\
         & \rule{0pt}{2.2ex} $10^4$ & 0.3986 & 0.4533 & 0.5033 & 0.3767 & 0.4319 & 0.4830 \\
         & \rule{0pt}{2.2ex} $10^5$ & 0.3952 & 0.4496 & 0.4995 & 0.3735 & 0.4285 & 0.4794 \\
        \hline
        Carr (2024) & \rule{0pt}{2.2ex} NA & 0.3948 & 0.4493 & 0.4994 & 0.3731 & 0.4283 & 0.4792 \\
        \hline
    \end{tabular}
    \caption{\textbf{Comparison to Carr's (2024) core-shell formulas.} Calculated values of $F_{\infty}$ obtained using the analytical expressions developed in the current paper (equations (\ref{EQ: Finf Final Result Fully - d=1})--(\ref{EQ: Finf Final Result Semi - d=3})) to those obtained using the core-shell analytical expressions developed previously by Carr (2024) (equations (29)--(34) in \cite{Carr:2024_FractionRelease}). Results are given for each combination of device geometry ($d=1,2,3$ for slab, cylinder, sphere) and coating permeability (fully-permeable, semi-permeable). Parameter values: $D(r)$ and $k(r)$ as per Figure \ref{FIG: Parameter Profiles - D and k} except with $k_{\text{min}} = 0\,\text{s}^{-1}$, $c_{0}(r) = 0.4[1-H(r-R/2)]\,\text{mol}\,\text{cm}^{-3}$, $R = 10^{-4}\,\text{cm}$, $P = 5\cdot 10^{-8}\,\text{cm}\,\text{s}^{-1}$, $N_{s} = 400$ [Current paper]; $D_{c} = 10^{-11}\,\text{cm}^2\,\text{s}^{-1}$, $D_{s} = 10^{-13}\,\text{cm}^2\,\text{s}^{-1}$, $k_{s} = 5\cdot 10^{-5}\,\text{s}^{-1}$, $c_{0}=0.4\,\text{mol}\,\text{cm}^{-3}$, $R = 10^{-4}\,\text{cm}$, $R_{c} = 0.5\cdot 10^{-4}\,\text{cm}$, $P = 5\cdot 10^{-8}\,\text{cm}\,\text{s}^{-1}$ [Carr (2024)].}
    \label{TAB: Coreshell Results Comparison}
\end{table}

\newpage
\section{Conclusions}
\label{SEC: Conclusions}
Recent research has highlighted the advantages of functionally-graded materials (FGMs) in drug delivery systems to achieve desired release profiles, and revealed how it is possible for drug particles to undergo a binding reaction and become irreversibly immobilised within the system. In this paper, we have presented novel deterministic-continuum and stochastic-discrete approaches for extracting insight into the drug release profile, $F(t)$, for functionally-graded delivery systems with binding reactions. 

The foundation of both approaches is a recently established reaction-diffusion model of drug release in the presence of binding reactions, where a first-order reaction term captures the binding reactions, and smooth spatially-varying diffusivity and reaction rate functions reflect the functionally-graded material \cite{Carr:2024_FunctionalRelease}. The deterministic-continuum approach, outlined in Section \ref{SEC: Analytical Approach}, yielded exact analytical expressions for calculating the total fraction of drug released, $F_{\infty}:= \lim_{t\rightarrow\infty}F(t)$. This was achieved by expressing $F_{\infty}$ in terms of the solution of a boundary value problem (derived from the governing reaction-diffusion model) and then solving analytically using a generalised eigenfunction expansion technique. The stochastic-discrete approach, outlined in Section \ref{SEC: Stochastic Approach}, developed a random walk model of the drug release process capturing variability in $F(t)$ and $F_{\infty}$. This was achieved by discretising the governing reaction-diffusion model in space and time to derive probabilities governing the movement, binding and release of individual drug particles. Simulation results presented in Section \ref{SEC: Results} showed that both  approaches are in good agreement with benchmark numerical approximations of $F_{\infty}$ and/or $F(t)$, and that the deterministic-continuum approach is able to reproduce values of $F_{\infty}$ obtained from the formulas derived in \cite{Carr:2024_FractionRelease} for the two limiting cases of the monolithic and core-shell system.

In summary, we have provided new analytical and computational tools for exploring the effect of system parameters (diffusivity and reaction rate functions, geometry of the device, coating permeability) on the drug release, insight which may be useful for designers and manufacturers of drug delivery systems. While both the deterministic-continuum and stochastic-discrete approaches are valid for general smooth diffusivity and reaction-rate functions, $D(r)$ and $k(r)$, it is important to note that they are both limited to the specific device configurations illustrated in Figure \ref{FIG: Device Geometries}, namely, radially-symmetric slab, cylinder and sphere geometries uniformly encapsulated in a thin coating that is either fully-permeable (zero resistance to drug release) or semi-permeable (finite resistance to drug release). Extending the deterministic-continuum and/or stochastic-discrete approaches to relax some of these assumptions, e.g. angular-dependent diffusivity and reaction-rate or non-uniform boundary conditions at the release surface, could be interesting and challenging directions for future research.

\section*{CRediT authorship contribution statement}
\textbf{Obi A. Carwood:} Writing -- original draft, Writing -- review \& editing, Formal analysis, Methodology, Validation, Visualization, Investigation, Software, Data Curation, Funding acquisition. \textbf{Elliot~J.~Carr:} Conceptualization, Writing -- review \& editing, Methodology, Supervision, Funding acquisition.

\section*{Data availability}
\label{sec:data_availability}
MATLAB code implementing all methods and reproducing the results of the paper is available on GitHub: \href{https://github.com/Obi-Carwood/Carwood2025}{https://github.com/Obi-Carwood/Carwood2025}.

\section*{Acknowledgements}
The first author completed this research as part of a Master of Philosophy (MPhil) project and acknowledges support provided by an Australian Government Research Training Program Scholarship.

% Endmatter
\appendix
\renewcommand{\thesection}{Appendix \Alph{section}}

\section{Derivation for $F_{\infty}$}
\label{APPENDIX: Derivation for Finf}
In this appendix, we derive the formula for $F_{\infty}$ given in equation (\ref{EQ: Finf - c}) for the radially-symmetric reaction-diffusion model (\ref{PDE: Governing Reaction-Diffusion})--(\ref{EQS: Governing BC2}). The working closely follows the working featured in \cite{Carr:2024_FractionRelease} for the simplified monolithic system (constant values of diffusivity, reaction-rate and initial concentration). By definition, $F_{\infty}$ is equal to the total cumulative amount of drug released from the device scaled by the initial amount of drug loaded into the device. Mathematically, this can be represented in terms of the concentration flux \cite{Carr:2024_FractionRelease}:
\begin{align}
\label{eq:Finf1_AppendixA}
    F_{\infty} &= \frac{\int_0^{\infty} \iint_{\omega} \left[ -D \nabla c \cdot \vec{n} \, \diff A \right] \diff t}{\iiint_{\Omega} c_0 \, \diff V},
\end{align}
where $\omega$ denotes the release surface(s) (see Figure \ref{FIG: Device Geometries}), $\Omega$ is the domain occupied by the device and $\vec{n}$ is the unit vector normal to $\omega$ directed outwards from $\Omega$. Under radial-symmetry, where $c$, $D$ and $c_{0}$ vary spatially in $r$ only, we have
\begin{align}
\label{eq:Finf2_AppendixA}
    F_{\infty} &= \frac{|\omega|\int_0^{\infty} -D(R) \frac{\partial c}{\partial r}(R,t) \, \diff t}{|\Psi| \int_0^R r^{d-1} c_0(r) \, \diff r},
\end{align}
where $|\omega|$ is the area of $\omega$ and $|\Psi|$ is the constant obtained by integrating out the non-radial coordinates in the denominator of (\ref{eq:Finf1_AppendixA}), with both quantities depending on the domain geometry ($d=1,2,3$ for the slab, cylinder, sphere):
\begin{align*}
    |\omega| =
    \begin{dcases}
        LW, \quad &\text{for $d=1$}, \\
        2\pi R H, \quad &\text{for $d=2$}, \\
        4\pi R^2, \quad &\text{for $d=3$}.
    \end{dcases}\qquad
    |\Psi| =
    \begin{dcases}
        LW, \quad &\text{for $d=1$}, \\
        2\pi H, \quad &\text{for $d=2$}, \\
        4\pi, \quad &\text{for $d=3$}.
    \end{dcases}
\end{align*}
Combining (\ref{eq:Finf2_AppendixA}) with the observation that $|\omega|/|\Psi| = R^{d-1}$ for all $d=1,2,3$ then yields the desired result~(\ref{EQ: Finf - c}):
\begin{align*}
    F_{\infty} &= \frac{R^{d-1}\int_0^{\infty} -D(R) \frac{\partial c}{\partial r}(R,t) \, \diff t}{\int_0^R r^{d-1} c_0(r) \, \diff r}.
\end{align*}

\section{New expression for $F_{\infty}$}
\label{APPENDIX: New expression for Finf}
In this appendix, we derive the new formula (\ref{EQ: Finf - C}) which expresses $F_{\infty}$ in terms of the solution, $C(r)$, of the boundary value problem (\ref{ODE: Transformed Reaction-Diffusion - C})--(\ref{EQ: Transformed BC2 - C}). The working closely follows the working featured in \cite{Carr:2024_FractionRelease} for the simplified monolithic system (constant values of diffusivity, reaction-rate and initial concentration). We begin by first multiplying the reaction-diffusion equation (\ref{PDE: Governing Reaction-Diffusion}) by $r^{d-1}$ and integrating over $r \in [0,R]$ to give:
\begin{align*}
    \int_0^R r^{d-1}\frac{\partial c}{\partial t} \, \diff r = \int_0^R \frac{\partial}{\partial r} \left( D(r) r^{d-1} \frac{\partial c}{\partial r} \right) \diff r - \int_0^R r^{d-1} k(r)c(r,t) \, \diff r.
\end{align*}
Next, reversing the order of the derivative and integral on the left-hand side and inserting the boundary conditions (\ref{EQS: Governing BC1}) on the right-hand side, we get,
\begin{align*}
    \frac{\diff}{\diff t} \int_0^R r^{d-1}c(r,t) \, \diff r = D(R) R^{d-1} \frac{\partial c}{\partial r}(R,t) - \int_0^R r^{d-1} k(r)c(r,t) \, \diff r.
\end{align*}
Integrating this equation over $t \in [0,\infty)$ then gives:
\begin{align*}
    \int_0^R r^{d-1} \left[ \lim_{t\to\infty} c(r,t) - c(r,0) \right] \diff r &= \int_0^{\infty} D(R) R^{d-1} \frac{\partial c}{\partial r}(R,t) \, \diff t - \int_0^R r^{d-1} k(r) \left(\int_0^{\infty} c(r,t) \, \diff t \right) \diff r.
\end{align*}
Noting the transformation $C(r)$ (\ref{EQ: Transformation c to C}), the initial condition (\ref{EQ: Governing IC}), and the fact that $c(r,t)$ approaches zero as $t$ tends to infinity, we get:
\begin{align*}
    \int_0^R -r^{d-1} c_0(r) \, \diff r &= \int_0^{\infty} D(R) R^{d-1} \frac{\partial c}{\partial r}(R,t) \, \diff t - \int_0^R r^{d-1} k(r) C(r) \, \diff r.
\end{align*}
Finally, rearranging and dividing by $\int_0^R r^{d-1} c_0(r) \, \diff r$ yields:
\begin{align*}
    \frac{R^{d-1} \int_0^{\infty} -D(R) \frac{\partial c}{\partial r}(R,t) \, \diff t}{\int_0^R r^{d-1} c_0(r) \, \diff r} &= 1 - \frac{\int_0^R r^{d-1} k(r) C(r) \, \diff r}{\int_0^R r^{d-1} c_0(r) \, \diff r}.
\end{align*}
Recall that the left-hand-side is precisely the formula for the total fraction of drug released ($F_{\infty}$) given in equation (\ref{EQ: Finf - c}) and hence we have derived the desired result (\ref{EQ: Finf - C}):
\begin{align*}
    F_{\infty} = 1 - \frac{\int_0^R r^{d-1} k(r) C(r) \,\text{d}r}{\int_0^R r^{d-1} c_0(r) \,\text{d}r}.
\end{align*}
We now turn our attention to deriving the boundary-value-problem (\ref{ODE: Transformed Reaction-Diffusion - C})--(\ref{EQ: Transformed BC2 - C}) satisfied by $C(r)$. Here, we begin by  integrating the governing reaction-diffusion equation (\ref{PDE: Governing Reaction-Diffusion})  over $t \in [0,\infty)$:
\begin{align*}
    \int_0^{\infty}\frac{\partial c}{\partial t}\, \diff t = \int_0^{\infty}\frac{1}{r^{d-1}} \frac{\partial}{\partial r} \left( D(r) r^{d-1} \frac{\partial c}{\partial r} \right) \diff t - \int_{0}^{\infty} k(r)c(r,t) \, \diff t,
\end{align*}
and then simplifying to yield:
\begin{align*}
    \lim_{t \to \infty} c(r,t)-c(r,0) = \frac{1}{r^{d-1}} \frac{\diff}{\diff r} \left( D(r) r^{d-1} \frac{\diff}{\diff r} \int_0^{\infty}c(r,t) \, \diff t \right) - k(r) \int_0^{\infty} c(r,t) \, \diff t.
\end{align*}
Again, noting the transformation $C(r)$ (\ref{EQ: Transformation c to C}), the initial condition (\ref{EQ: Governing IC}), and the fact that $c(r,t)$ approaches zero as $t$ tends to infinity, we get the differential equation satisfied by $C(r)$ as stated in equation (\ref{ODE: Transformed Reaction-Diffusion - C}), namely:
\begin{align*}
    -c_0(r) = \frac{1}{r^{d-1}} \frac{\diff}{\diff r} \left( D(r) r^{d-1} \frac{\diff C}{\diff r} \right) - k(r) C(r).
\end{align*}
Lastly, taking a similar approach of integrating the governing boundary conditions (\ref{EQS: Governing BC1})--(\ref{EQS: Governing BC2}) over $t \in [0,\infty)$ yields the boundary conditions (\ref{EQ: Transformed BC1 - C})--(\ref{EQ: Transformed BC2 - C}) satisfied by $C(r)$ (see \cite{Carr:2024_FractionRelease} for further detail).

\section{Finite volume method}
\label{APPENDIX: Finite volume method}
In this appendix, we discretise the reaction-diffusion model (\ref{PDE: Governing Reaction-Diffusion})--(\ref{EQS: Governing BC2}) in space to derive the system of ODEs given in equation (\ref{ODE: Finite Volume - Markov Interpretation}) for the semi-permeable coating problem. Spatial discretisation is carried out using a standard vertex-centred finite volume method on a uniform spatial grid consisting of $M$ nodes, located at $r_{i} = (i-1)h$ for $i = 1,\hdots,M$ where $h = R/(M-1)$, and $M$ control volumes spanning the intervals $r\in[w_{i},e_{i}]$ for $i = 1,\hdots,M$, where $w_{i} = \max\{0,(i-3/2)h\}$ and $e_{i} = \min\{(i-1/2)h,R\}$ are the ``west'' and ``east'' boundaries of control volume $i$, respectively. Multiplying (\ref{PDE: Governing Reaction-Diffusion}) by $r^{d-1}$ and then integrating over $[w_{i},e_{i}]$ yields:
\begin{align*}
\frac{\diff \overline{c}_{i}}{\diff t} = \frac{D(e_{i})e_{i}^{d-1}}{V_{i}}\frac{\partial c}{\partial r}(e_{i},t) - \frac{D(w_{i})w_{i}^{d-1}}{V_{i}}\frac{\partial c}{\partial r}(w_{i},t) - \overline{kc}_{i},
\end{align*}
where $\overline{c}_{i}$ and $\overline{kc}_{i}$ are, respectively, the averaged values of $c(r,t)$ and $k(r)c(r,t)$ over control volume $i$:
\begin{align*}
\overline{c}_{i} = \frac{1}{V_{i}}\int_{w_{i}}^{e_{i}}r^{d-1}c(r,t) \, \diff r,\quad \overline{kc}_{i} = \frac{1}{V_{i}}\int_{w_{i}}^{e_{i}}r^{d-1}k(r)c(r,t) \, \diff r,\quad V_{i} = \int_{w_{i}}^{e_{i}} r^{d-1}\,\diff r,
\end{align*}
with $c_{i}$ denoting the approximation to $c(r_{i},t)$. Using the standard finite volume approach of approximating averaged values over control volumes by the values at the corresponding nodes then gives:
\begin{align*}
\frac{\diff c_{i}}{\diff t} = \frac{D(e_{i})e_{i}^{d-1}}{V_{i}}\frac{\partial c}{\partial r}(e_{i},t) - \frac{D(w_{i})w_{i}^{d-1}}{V_{i}}\frac{\partial c}{\partial r}(w_{i},t) - k(r_{i})c_{i} .
\end{align*}
Finally, inserting the boundary conditions (\ref{EQS: Governing BC1}) and (\ref{EQS: Governing BC2}) (semi-permeable case only), which apply at $r = w_{1} = 0$ and $r = e_{N} = R$, respectively, and using second-order central difference approximations for the remaining derivatives yields the following spatially-discrete system of ODEs:
\begin{align*}
\frac{\diff c_{1}}{\diff t} &= - \frac{D(e_{1})e_{1}^{d-1}}{V_{1}h}c_{1} +\frac{D(e_{1})e_{1}^{d-1}}{V_{1}h}c_{2} - k(r_{1})c_{1},\\
\frac{\diff c_{i}}{\diff t} &=  \frac{D(w_{i})w_{i}^{d-1}}{V_{i}h}c_{i-1} - \left[\frac{D(w_{i})w_{i}^{d-1}}{V_{i}h} + \frac{D(e_{i})e_{i}^{d-1}}{V_{i}h}\right]c_{i} + \frac{D(e_{i})e_{i}^{d-1}}{V_{i}h}c_{i+1} - k(r_{i})c_{i},\quad \text{for $i=2,\hdots,M-1$},\\
\frac{\diff c_{M}}{\diff t} &= \frac{D(w_{M})w_{M}^{d-1}}{V_{M}h}c_{M-1} - \frac{D(w_{M})w_{M}^{d-1}}{V_{M}h}c_{M}  - k(r_{M})c_{M} - \frac{PR^{d-1}}{V_{M}}c_{M}.
\end{align*}
The above system can then be expressed in the form of the ODE system (\ref{ODE: Finite Volume - Markov Interpretation}), where coefficients of $c_{1},\hdots,c_{M}$ involving the diffusivity $D(r)$, reaction-rate $k(r)$ and mass transfer coefficient $P$ appear in $\vec{A}_\vec{m}$, $\vec{A}_\vec{b}$ and $\vec{A}_\vec{r}$, respectively. In summary, the non-zero entries $\vec{A}_\vec{m}$, $\vec{A}_\vec{b}$ and $\vec{A}_\vec{r}$ are given by:
\begin{gather*}
a^{m}_{1,1} = - \frac{D(e_{1})e_{1}^{d-1}}{V_{1}h},\quad a^{m}_{1,2} = \frac{D(e_{1})e_{1}^{d-1}}{V_{1}h},\\
a^{m}_{i,i-1} = \frac{D(w_{i})w_{i}^{d-1}}{V_{i}h},\quad a^{m}_{i,i} = -\left[\frac{D(w_{i})w_{i}^{d-1}}{V_{i}h} + \frac{D(e_{i})e_{i}^{d-1}}{V_{i}h}\right],\quad  a^{m}_{i,i+1} = \frac{D(e_{i})e_{i}^{d-1}}{V_{i}h},\quad \text{for $i=2,\hdots,M-1$},\\
a^{m}_{M,M-1} = \frac{D(w_{M})w_{M}^{d-1}}{V_{M}h},\quad a^{m}_{M,M} = - \frac{D(w_{M})w_{M}^{d-1}}{V_{M}h},\\
a^{b}_{i,i} = k(r_{i}),\quad \text{for $i=1,\hdots,M$},\quad a^{r}_{M,M} = \frac{PR^{d-1}}{V_{M}},
\end{gather*}
where $a^{m}_{i,j}$, $a^{b}_{i,j}$ and $a^{r}_{i,j}$ denote the entries in row $i$ and column $j$ of $\mathbf{A}_\vec{m}$, $\vec{A}_\vec{b}$ and $\vec{A}_\vec{r}$, respectively. 

\section{Numerical computation of $F_{\infty}$ and $F(t)$}
\label{APPENDIX: Numerical Computation}
In section \ref{SEC: Verification}, numerical benchmark approximations are used to validate our deterministic-continuum and stochastic-discrete approaches for calculating $F_{\infty}$ and $F(t)$. In this appendix, we explain how these benchmark approximations are computed by combining a numerical solution of the governing reaction-diffusion model (\ref{PDE: Governing Reaction-Diffusion})--(\ref{EQS: Governing BC2}) with the following analytical expression for $F(t)$:
\begin{align}
    \label{EQ: Ft - c}
F(t) = \frac{R^{d-1} \int_0^{t} -D(R) \frac{\partial c}{\partial r}(R,s) \, \diff s}{\int_0^R r^{d-1} c_0(r) \, \diff r}.
\end{align}
We solve the reaction-diffusion model (\ref{PDE: Governing Reaction-Diffusion})--(\ref{EQS: Governing BC2}) numerically by discretising in space using the vertex-centered finite volume method outlined in \ref{APPENDIX: Finite volume method} and in time using the backward Euler method. As for the stochastic-discrete approach (Section \ref{SEC: Stochastic Approach}), discretisation is carried out using uniform grids in both space and time, defined by $r_{i} = (i-1)h$ for $i = 1,\hdots,M$ and $t_{n} = n\tau$ for $n = 1,\hdots,N_t$, where $h = R/(M-1)$ and $\tau = T/N_t$ (however, we assume $M$ and $N_t$ may take on different values to those featuring in the stochastic-discrete approach). For each time $t=t_n$, we then compute the fraction of drug released, $F(t_n)$, by processing the following steps: (i) evaluate $F(t)$ (\ref{EQ: Ft - c}) at $t = t_{n}$ (ii) approximate the spatial derivative, $\frac{\partial c}{\partial r}(R,s)$, using a backward-difference approximation (iii) numerically compute both the numerator and denominator integrals featuring in $F(t)$ (\ref{EQ: Ft - c}) using in-built MATLAB functions. Finally, we take $F_{\infty}$ to be $F(T)$ (i.e. $F(t_{N_{t}})$), making sure to choose $T$ large enough to capture the limiting behaviour of $F(t)$. For full details on the computation of the numerical benchmarks, the interested reader is referred to our MATLAB code available on GitHub; see \nameref{sec:data_availability}.

% References
\bibliographystyle{unsrt}
\bibliography{References}

\end{document}